\documentclass[twocolumn]{aastex701}
\usepackage{chngcntr}

\counterwithout{table}{section}

\renewcommand{\thetable}{\arabic{table}}

\usepackage{gensymb}
\usepackage{xcolor}
\usepackage{booktabs}
\usepackage{amsmath}   
\usepackage{array} 
\usepackage{tabularx}
\usepackage{array}
\usepackage{multirow}

\begin{document}

\title{Hot, Photoionized X-ray Gas in Two Luminous Type 2 Quasars: Chandra-HST Evidence for a Wind-Driven Sequence}

\author[orcid=0000-0001-8112-3464]{Anna~Trindade~Falcão}
\affiliation{NASA Goddard Space Flight Center, Code 662, Greenbelt, MD 20771, USA}
\email{annatrindadefalcao@gmail.com}

\author[orcid=0000-0003-4073-8977]{S. Kraemer}
\affiliation{The Catholic University of America, Washington, DC 20064, USA}
\email{kraemer@cua.edu}

\author[orcid=0000-0002-5718-2402]{L. Feuillet}
\affiliation{The Catholic University of America, Washington, DC 20064, USA}
\email{feuilletl@cua.edu}

\author[orcid=0000-0001-9815-9092]{R. Middei}
\affiliation{Harvard-Smithsonian Center for Astrophysics, 60 Garden St., Cambridge, MA 02138, USA}
\affiliation{INAF Osservatorio Astronomico di Roma, Via Frascati 33, 00078 Monte Porzio Catone, RM, Italy}
\email{riccardo.middei@inaf.it}

\author[orcid=0000-0003-2971-1722]{T. J. Turner}
\affiliation{Eureka Scientific, Inc., 2452 Delmer Street, Suite 100, Oakland, CA 94602, USA}
\email{turnertjane@gmail.com}

\author[orcid=0000-0002-0982-0561]{J. Reeves}
\affiliation{The Catholic University of America, Washington, DC 20064, USA}
\affiliation{INAF - Osservatorio Astronomico di Brera, Via Bianchi 46, 23807, Merate (LC), Italy}
\email{james.n.reeves456@gmail.com}

\author[orcid=0000-0002-2629-4989]{V. Braito}
\affiliation{The Catholic University of America, Washington, DC 20064, USA}
\affiliation{INAF - Osservatorio Astronomico di Brera, Via Bianchi 46, 23807, Merate (LC), Italy}
\affiliation{Dipartimento di Fisica, Università di Trento, Via Sommarive 14, Trento 38123, Italy}
\email{braito@cua.edu}

\author[orcid=0000-0001-5655-1440]{A. Ptak}
\affiliation{NASA Goddard Space Flight Center, Code 662, Greenbelt, MD 20771, USA}
\email{andrew.ptak@nasa.gov}

\author[orcid=0000-0003-2450-3246]{H. R. Schmitt}
\affiliation{Naval Research Laboratory, Washington, DC 20375, USA}
\email{schmitt.henrique@gmail.com}

\author[orcid=0000-0002-3365-8875]{T. C. Fischer}
\affiliation{AURA for ESA, Space Telescope Science Institute, 3700 San Martin Drive, Baltimore, MD 21218, USA}
\email{tfischer@stsci.edu}

\author[orcid=0000-0002-6465-3639]{D. M. Crenshaw}
\affiliation{Georgia State University, 25 Park Place, Suite 600, Atlanta, GA 30303, USA}
\email{dmichaelcrenshaw@gmail.com}

\author[orcid=0000-0001-6947-5846]{Luis C. Ho}
\affiliation{Kavli Institute for Astronomy and Astrophysics, Peking University,
Beijing 100871, China}
\affiliation{Department of Astronomy, School of Physics, Peking University,
Beijing 100871, China}
\email{lho.pku@gmail.com}

\author[orcid=0000-0002-4917-7873]{M. Revalski}
\affiliation{Space Telescope Science Institute, 3700 San Martin Drive, Baltimore, MD 21218, USA}
\email{mrevalski@stsci.edu}

\author[orcid=0000-0003-1772-0023]{T. Storchi-Bergmann}
\affiliation{Universidade Federal do Rio Grande do Sul, IF, CEP 15051, 91501-970 Porto Alegre, RS, Brazil}
\email{thaisa.storchi.bergmann@gmail.com}

\author[orcid=0000-0001-9191-9837]{M. Vestergaard}
\affiliation{DARK, Niels Bohr Institute, University of Copenhagen, Jagtvej 155A, 2200 Copenhagen N, Denmark}
\affiliation{Department of Astronomy and Steward Observatory, 933 Cherry Avenue, Tucson, AZ 85721, USA}
\email{mvester@nbi.ku.dk}

\author[orcid=0000-0003-4888-2009]{C. M. Gaskell}
\affiliation{Department of Astronomy and Astrophysics, University of California, Santa Cruz, CA 95064, USA}
\email{mgaskell@ucsc.edu}

\author[orcid=0000-0002-2203-7889]{W. P. Maksym}
\affiliation{NASA Marshall Space Flight Center, Huntsville, AL 35812, USA}
\email{walter.p.maksym@nasa.gov}

\author[orcid=0000-0001-5060-1398]{M. Elvis}
\affiliation{Harvard-Smithsonian Center for Astrophysics, 60 Garden St., Cambridge, MA 02138, USA}
\email{melvis@cfa.harvard.edu}

\author[orcid=0000-0003-1810-0889]{M. J. Ward}
\affiliation{Centre for Extragalactic Astronomy, Department of Physics, University of Durham, South Road, Durham DH1 3LE, UK}
\email{martin.ward@durham.ac.uk}

\author[orcid=0000-0002-6766-0260]{H. Netzer}
\affiliation{School of Physics and Astronomy, Tel Aviv University, Tel Aviv 69978, Israel}
\email{hagainetzer@gmail.com}
 
\begin{abstract}
We present new Chandra/ACIS-S imaging spectroscopy of two luminous type~2 quasars, FIRST~J120041.4+314745 ($z$=0.116) and 2MASX~J13003807+5454367 ($z$=0.088), and compare their X-ray emission with Hubble Space Telescope [O~III]$\lambda$5007 morphologies and kinematics. Both systems show kiloparsec-scale soft X-ray emission. In FIRST~J120041, the X-ray morphology is clumpy and closely follows the [O~III] structures, with surface-brightness peaks co-spatial with the highest [O~III] velocities (600-750~km~s$^{-1}$) and broadest line widths ($\sim$1700~km~s$^{-1}$). In 2MASX~J130038, the X-ray emission is centrally concentrated and weakly correlated with rotational [O~III] kinematics. Spectral modeling indicates that photoionization dominates the soft X-rays in both quasars. The inferred hot-gas reservoirs are substantial, $M_{\rm x-ray}\sim4.5\times10^{8}M_\odot$ (FIRST~J120041) and $M_{\rm x-ray}\sim1.8\times10^{8}M_\odot$ (2MASX~J130038), exceeding the outflowing [O~III] masses by factors of $\sim$4 and $\sim$16. In 2MASX~J130038, we identify a tentative blueshifted Fe~XXVI~Ly$\alpha$ line at $E_{\rm rest}=7.14\pm0.06$~keV ($v\sim7600$~km~s$^{-1}$), consistent with a hot wind confined to the inner few hundred parsecs. Combining these results with a broader sample of twelve type~2 quasars, we argue that luminous quasars evolve along a continuous feedback sequence regulated by progressive clearing of circumnuclear gas. As AGN radiation and winds pierce the surrounding medium, systems transition from heavily enshrouded, compact configurations to phases where the X-ray and [O~III] components strongly couple and, eventually, to energetically dominant outflows. FIRST~J120041 and 2MASX~J130038 represent two points along this sequence, tracing the emergence and growth of hot winds as primary drivers of quasar-scale feedback.
\end{abstract}


\section{Introduction} 
\label{sec:intro}
Active galactic nuclei (AGNs) have the potential to influence the evolution of galaxies through mass outflows that interact with the surrounding interstellar medium (ISM). These feedback processes may regulate the co-evolution of supermassive black holes (SMBHs) and their host galaxies, helping to establish observed scaling relations such as the $M_{\rm BH}$-$\sigma_*$ correlation between SMBH mass and bulge velocity dispersion \citep[e.g.,][]{gebhardt_relationship_2000, begelman_agn_2004}. Despite their potential significance, the physical mechanisms that might drive AGN feedback remain under debate, largely because they are difficult to isolate observationally. The relevant processes operate on different spatial and temporal scales, and many potential mechanisms, such as induction heating, shocks, or outflows \citep[e.g.,][]{gaspari_unifying_2017, gaspari_linking_2020, trindade_falcao_hubble_2021, falcone_analysis_2024}, leave ambiguous or indirect signatures that make it challenging to identify the dominant driver in any given system.

At low luminosities (bolometric luminosities $L_{\rm bol}<10^{45}~\text{erg~s}^{-1}$), nearby ($z<0.1$) Seyfert galaxies provide valuable insight through narrow-line region (NLR) outflows. Optical spectroscopy of the [O~III]$\lambda5007$ line has been central to mapping NLR kinematics, revealing signatures of AGN-driven winds launched from near the SMBH as well as outflows arising from \textit{in situ} gas entrained within the host galaxy disk \citep{fischer_determining_2013, crenshaw_feedback_2015, fischer_gemini_2017}. Radiative acceleration is often invoked with considerable success to explain these outflows \citep{meena_radiative_2021, meena_investigating_2023, falcone_analysis_2024, tutterow_shape_2025}, yet the dynamics and origin of NLR winds are still not fully understood \citep[e.g.,][]{gaspari_unifying_2017, gaspari_linking_2020}.

At higher luminosities, type~2 quasars (QSO2s) provide a unique window into feedback operating on kiloparsec scales. Using Hubble Space Telescope (HST) [O~III] imaging and Space Telescope Imaging Spectrograph (STIS) spectroscopy, \citet{fischer_hubble_2018} showed that luminous, local QSO2s ($L_{\rm bol}\gtrsim10^{45.8}~\text{erg~s}^{-1}$, $z\lesssim0.12$) can host NLRs extending over several kiloparsecs \citep[see also e.g.,][]{storchi-bergmann_bipolar_2018}, with clear outflow signatures in their inner regions  \citep[see also e.g.,][]{de_oliveira_gauging_2021}. However, they also found that not all extended [O~III] NLRs are accompanied by equally extended [O~III] outflows. This discrepancy raises a key question: if outflows are radiatively accelerated and the AGN is luminous enough to ionize gas across kiloparsec scales, why do some outflows appear to stall near the nucleus?

One possibility is that hot X-ray winds provide the dominant dynamical driver, entraining and accelerating cooler gas phases such as the [O~III]-emitting component \citep{trindade_falcao_hubble_2021}. Indeed, ultra-fast outflows (UFOs) detected in X-rays have been linked to galaxy-scale molecular winds in ultra-luminous infrared galaxies (ULIRGs; \citealt{tombesi_wind_2015}), suggesting a physical connection between hot winds and cooler outflowing gas. In nearby Seyferts, Chandra imaging reveal extended soft X-ray emission that is spatially coincident with [O~III] structures \citep[e.g.,][]{young_chandra_2001, wang_deep_2011, maksym_cheers_2019}, while blueshifted X-ray emission lines show velocities comparable to those observed in the optical gas \citep[e.g.,][]{kraemer_physical_2015, kallman_census_2014}. In some systems, the X-ray outflow extends even beyond the optical NLR, implying that the hot phase may dominate the total energy budget \citep{kraemer_mass_2020}.

These correlations persist at higher luminosities. The luminous QSO2 Mrk~34, for example, shows a well-defined [O~III] bicone and high-velocity outflows ($\sim1000~\text{km~s}^{-1}$) extending to $\sim500$~pc \citep{fischer_hubble_2018}. Multi-component modeling of its NLR \citep{revalski_quantifying_2018, revalski_erratum_2019} combined with spatially resolved STIS spectroscopy yielded a peak outflow rate $\dot M\sim10.3~M_{\odot}~\text{yr}^{-1}$ at $\sim1.3$~kpc and peak kinetic luminosity $\dot E_{k}\sim1.3\times10^{42}~\text{erg~s}^{-1}$ at $\sim480$~pc \citep{trindade_falcao_hubble_2021-1}. The close morphological agreement between [O~III] and soft X-ray emission in Mrk~34 reinforces the idea of a physical coupling between these phases.

To investigate this connection further, we obtained new Chandra/ACIS-S observations of two luminous QSO2s from the \citet{fischer_hubble_2018} sample: FIRST~J120041.4+314745 ($z=0.116$, 2.04~kpc~arcsec$^{-1}$; hereafter FIRST~J120041) and 2MASX~J13003807+5454367 ($z=0.088$, 1.59~kpc~arcsec$^{-1}$; hereafter 2MASX~J130038). These objects are the two [O~III]$\lambda$5007-brightest ($\log{L_{\rm [OIII]}/{\rm erg~s}^{-1}}=43.1$ and 42.7, respectively; \citealt{trindade_falcao_hubble_2021-1}) of the extended sources following Mrk~34 and have the largest deprojected NLRs ($R_{\rm [O~III]}=6.1$~kpc and 4.7~kpc, respectively; \citealt{fischer_hubble_2018}). Despite their similar NLR sizes, their [O~III] outflows show strikingly different kinematics: FIRST~J120041 hosts high-velocity outflows extending to $\sim$1~kpc, whereas 2MASX~J130038 shows predominantly rotational motion. Photoionization modeling of their STIS spectra underscores this contrast, with FIRST~J120041 showing [O~III] mass outflow rates and kinetic luminosities that exceed those of 2MASX~J130038 by factors of $\sim$210 and $\sim$100, respectively \citep{trindade_falcao_hubble_2021-1}.

These contrasting properties make the two quasars an ideal testbed for isolating the role of hot, X-ray-emitting gas in driving large-scale ionized outflows. In this work, we combine the sub-arcsecond resolution of Chandra with archival HST imaging and spectroscopy to (1) map the spatially resolved ionization structure of the X-ray-emitting gas, and (2) test for morphological and kinematic correspondences between the X-ray and optical phases. By comparing two quasars with similarly extended optical NLRs but drastically different outflow kinematics and energetics, we aim to assess the physical role of hot winds in AGN feedback and to place these systems within a broader evolutionary sequence of quasar activity.

\section{Observations and Data Reduction} 
\label{sec:obs_data_reduction}

This study is based on new Chandra/ACIS-S observations of the QSO2s FIRST~J120041 and 2MASX~J130038, each with a total effective exposure time of $\sim$90~ks (Table~\ref{tab:observations}). All datasets are publicly available through the Chandra Data Collection (CDC) ``332" (\dataset[doi:10.25574/cdc.332]{https://doi.org/10.25574/cdc.332}).

\begin{table}
\begin{center}
\caption{Chandra/ACIS-S observations of the two QSO2s analyzed in this work. Columns list the observation ID, date, effective exposure time, principal investigator (PI), and astrometric offsets (dx, dy) applied to align each dataset (see Section~\ref{sec:obs_data_reduction}).}
\label{tab:observations}
\begin{tabular}{ccccc}
\hline
ObsID & Date & Time & PI & Offset \\
& & (ks) & & (dx, dy)\\
\hline
\multicolumn{5}{c}{\textbf{2MASX~J13003807+5454367}} \\
26787 & 2024-01-22 & 14.21 & Kraemer & -0.360, -0.566 \\
27160 & 2023-02-09 & 30.20 & Kraemer & -1.427, -1.314 \\
27161 & 2024-05-16 & 42.02 & Kraemer & +0.586, -0.238 \\
\hline
\multicolumn{5}{c}{\textbf{FIRST~J120041.4+314745}} \\
26786 & 2024-01-10 & 57.32 & Kraemer & +1.009, -0.189 \\
27158 & 2024-05-24 & 25.74 & Kraemer & +1.022, -1.382 \\
\hline
\end{tabular}
\tablecomments{dx and dy are given in units of sky pixels}
\end{center}
\end{table}

We retrieved the raw event files from the Chandra Data Archive\footnote{\url{https://cda.harvard.edu/chaser/}} and reprocessed them with \texttt{CIAO}~4.17\footnote{\url{http://cxc.harvard.edu/ciao}} \citep{fruscione_ciao_2006}, applying the latest calibration files through the \texttt{chandra\_repro}\footnote{\url{https://cxc.cfa.harvard.edu/ciao/ahelp/chandra\_repro.html}} script. Subpixel event repositioning was enabled to fully exploit the angular resolution of Chandra and resolve extended structures on sub-arcsecond scales. 
 
To ensure precise registration among individual observations, we refined the astrometry using the \texttt{fine\_astro}\footnote{\url{https://cxc.cfa.harvard.edu/ciao/ahelp/fine\_astro.html}} routine, which cross-matches detected X-ray sources with an external reference catalog and applies small linear shifts to the event files. The resulting positional offsets (dx, dy) in units of sky pixels are listed in Table~\ref{tab:observations} and are consistent with the nominal uncertainties of the Chandra aspect solution. The reprocessed and aligned event files were then merged to produce the final datasets used for the imaging analysis described in Section~\ref{sec:morphology}.

Spectra were extracted using the \texttt{specextract}\footnote{\url{https://cxc.cfa.harvard.edu/ciao/ahelp/specextract.html}} task. For FIRST~J120041, we used a circular aperture of radius $3.5''$ centered at R.A.=12:00:41.4456, decl.=+31:47:46.062, while for 2MASX~J130038 we adopted a $2.5''$ aperture centered at R.A.=13:00:38.2106, decl.=+54:54:36.739. These apertures encompass the full extent of the detected emission. Background spectra were extracted from nearby, source-free regions with radii of $10''$ (centered at R.A.=12:00:43.3043, decl.=+31:48:05.418 for FIRST~J120041, and R.A.=13:00:41.4230, decl.=+54:54:53.053 for 2MASX~J130038).  

The individual spectra and corresponding response files were combined using \texttt{combine\_spectra}\footnote{\url{https://cxc.cfa.harvard.edu/ciao/ahelp/combine\_spectra.html}}, and the resulting merged spectra were binned to a minimum of 10 counts per bin. All spectral fitting was performed in the rest-frame 0.3-8~keV energy range.

\section{Spectral Properties of the X-ray Emission}
\label{sec:spec_properties}

\subsection{Phenomenological Models}
\label{sec:pheno}

We first model the spectra using phenomenological, multi-component fits implemented in \texttt{Sherpa}\footnote{\url{https://cxc.cfa.harvard.edu/sherpa/}}. These models are intended to identify the main continuum components and locate prominent emission features without imposing strong physical assumptions \citep[e.g.,][]{trindade_falcao_deep_2023}.

The baseline model consists of an absorption component (\texttt{xstbabs}) allowed to vary above the Galactic value ($N_{\rm H}= 1.66\times10^{20}$~cm$^{-2}$ for FIRST~J120041, and $N_{\rm H}= 1.74\times10^{20}$~cm$^{-2}$ for 2MASX~J130038, calculated using the \texttt{HEASARC} tool\footnote{\url{https://heasarc.gsfc.nasa.gov/cgi-bin/Tools/w3nh/w3nh.pl}}, \citealt{bekhti_hi4pi_2016}), combined with a series of redshifted Gaussian lines (\texttt{xszgauss}) modeling discrete emission features. The Gaussian components have line widths ($\sigma$) free to vary, but are required to satisfy $\sigma\geq$0.1~keV, corresponding to the ACIS-S energy resolution.

When required by the data, we include an additional hard X-ray continuum component at $E\gtrsim$3~keV to account for nuclear AGN emission. In cases where the absorbed nuclear continuum is sufficiently strong to dominate the hard-band spectrum over the diffuse, non-nuclear emission (as in 2MASX~J130038), this component is modeled using the physically motivated torus reflection model \texttt{MYTorus} \citep{murphy_x-ray_2009}. In the \texttt{MYTorus} configuration, the scattered continuum (\texttt{myts}) and fluorescent line (\texttt{mytl}) components are linked to a common column density, with the inclination fixed at 90$^\circ$.

In contrast, for sources in which the nuclear emission does not dominate the total X-ray output but is still required to reproduce the observed continuum at $E\gtrsim$3~keV (as in FIRST~J120041), we adopt a simpler absorbed power-law component (\texttt{xspowerlaw}). Owing to limited photon statistics, the photon index is fixed at $\Gamma = 1.9$ for all \texttt{MYTorus} and \texttt{xspowerlaw} components.
 
The phenomenological models were constructed iteratively, beginning with a minimal absorbed continuum and adding components only when statistically and visually warranted by the data. After fitting the baseline model, we introduced Gaussian emission lines one at a time, retaining them only if they produced a significant improvement in the fit statistic and removed coherent residual features.

Because we adopt the Cash statistic (Cstat) for low-count spectra, we evaluate model improvement using the change in C-statistic ($\Delta$Cstat) together with the removal of localized residuals. Gaussian components are retained only when they reduce the C-statistic by $\Delta$Cstat $\gtrsim 9$ for two additional free parameters and when the associated residuals are repeatable across independent observations.

For nested models, $\Delta$Cstat asymptotically follows a $\chi^{2}$ distribution under Wilks’ theorem; for two additional degrees of freedom, $\Delta$Cstat $\approx 9$ corresponds to an improvement at approximately the 99\% confidence level. We therefore adopt $\Delta$Cstat $\gtrsim 9$ as a conservative empirical threshold for model inclusion throughout this work, including the physically motivated models in Section~\ref{sec:physical}. This criterion is intended for model selection rather than as a formal line-detection significance test, which would require dedicated Monte Carlo simulations \citep[e.g.,][]{kaastra_use_2017}.

The phenomenological fits are used to identify statistically significant spectral structure and to guide the construction of the physically motivated models presented in Section~\ref{sec:physical}. Given the limited spectral resolution and photon statistics of the ACIS-S data, the Gaussian components should be interpreted as representations of blended line complexes rather than formal detections of individual atomic transitions.
 
\begin{figure*}
  \centering
  \begin{minipage}[t]{0.49\textwidth}
    \centering
    \includegraphics[width=\linewidth]{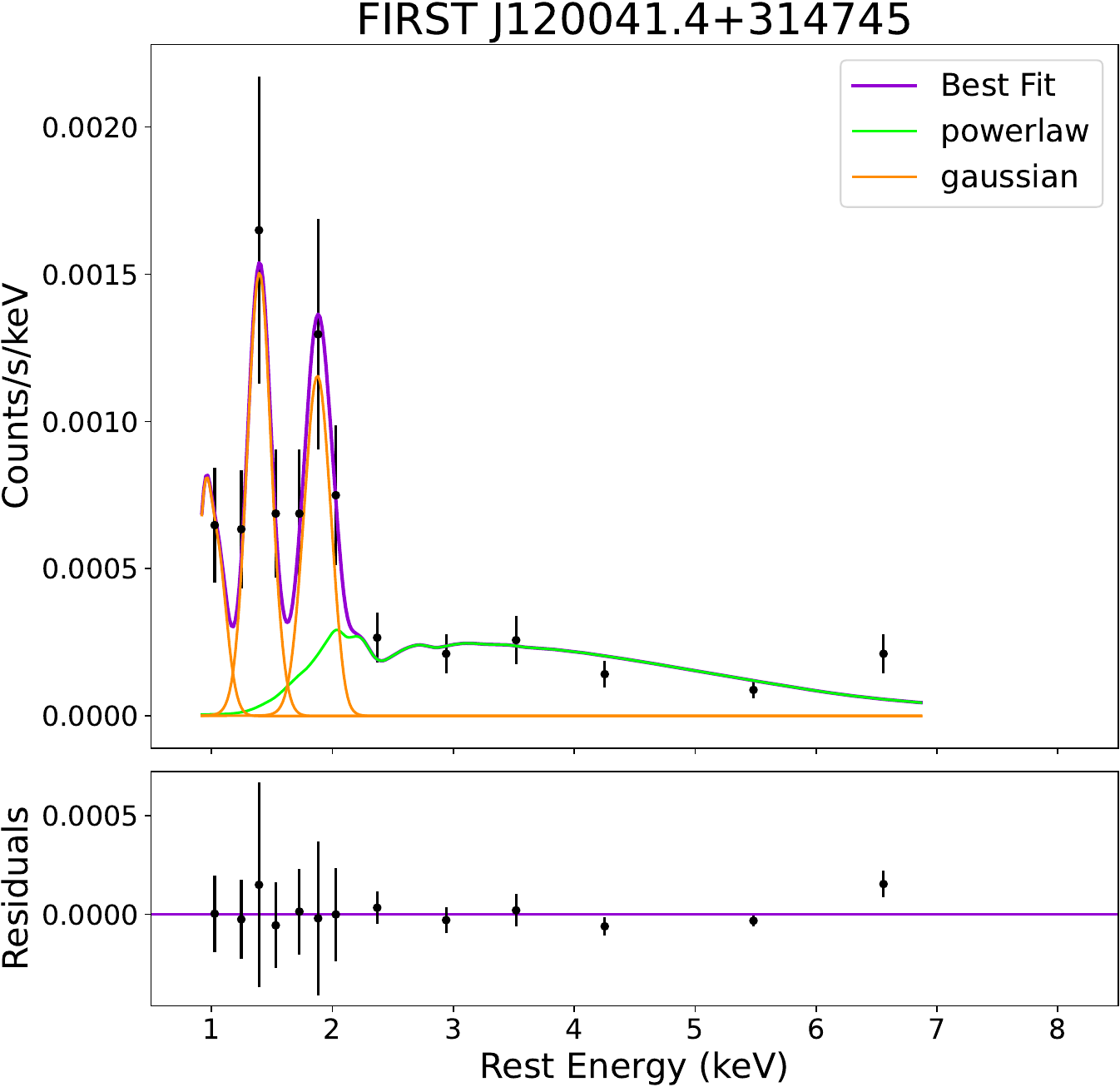}
  \end{minipage}%
  \hfill
  \begin{minipage}[t]{0.49\textwidth}
    \centering
    \includegraphics[width=\linewidth]{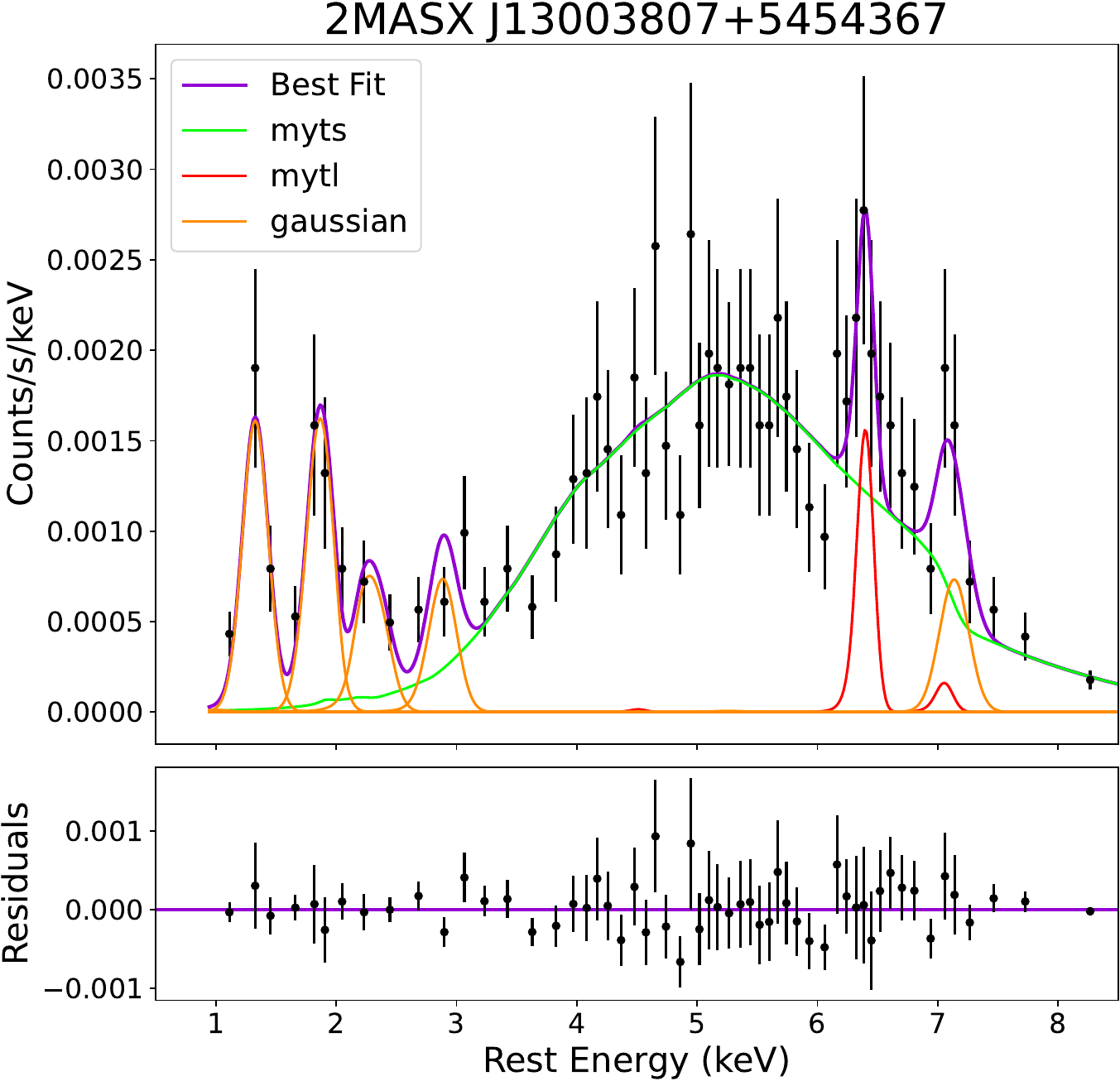}
  \end{minipage}
\caption{Rest-frame 0.3-8~keV spectra of FIRST~J120041 (left) and 2MASX~J130038 (right). Top panels show the best-fit phenomenological models; bottom panels show residuals relative to the fits.}
 \label{fig:pheno} 
 \end{figure*}

Figure~\ref{fig:pheno} shows the best-fit phenomenological models, and the resulting parameters are summarized in Table~\ref{tab:pheno}.

\begin{table*}
\centering
\caption{Best-fit parameters from phenomenological spectral models. Fits include an absorption component, Gaussian emission lines, and, where required, either a simple power law or the \texttt{MYTorus} reprocessing model. Quoted errors are 1$\sigma$.}
\label{tab:pheno}
\begin{tabular}{lllcc}
\toprule
Component & Physical Interpretation & Parameter & Value & Identification \\
\midrule
\multicolumn{5}{c}{\textbf{FIRST~J120041.4+314745}} \\
\midrule
\multicolumn{5}{c}{Phenomenological fit (Cstat/d.o.f. = 2.92)} \\
\hline
\texttt{xstbabs} & Foreground absorption & $N_{\rm H}$ & $(2.69^{+0.94}_{-0.96}) \times10^{22}$ & \\
\texttt{xszgauss} & Emission Line &$E_{\rm rest}$ & $0.75^{+0.15}_{-0.54}$ & O~VIII (blended) \\
 & & norm & $(1.03^{+277.32}_{-1.02})$ & \\
 \texttt{xszgauss} & Emission Line & $E_{\rm rest}$ & $1.30^{+0.04}_{-0.05}$ & Mg~XI (blended) \\
 & & norm & $(1.40^{+3.71}_{-1.12}) \times 10^{-4}$ & \\
\texttt{xszgauss} & Emission Line & $E_{\rm rest}$ & $1.84^{+0.04}_{-0.04}$ & Si~XIII (blended)\\
 & & norm & $(5.67^{+7.50}_{-3.45}) \times 10^{-6}$ & \\
\texttt{xspowerlaw} & AGN Continuum & norm & $(8.31^{+1.73}_{-1.51}) \times 10^{-6}$ & \\
\midrule
\multicolumn{5}{c}{\textbf{2MASX~J13003807+5454367}} \\
\midrule
\multicolumn{4}{c}{Phenomenological fit (Cstat/d.o.f. = 0.76)} \\
\hline
\texttt{xstbabs} & Foreground absorption & $N_{\rm H}$ & $<3.50\times10^{22}$ & \\
\texttt{xszgauss} & Emission Line & $E_{\rm rest}$ & $1.30^{+0.02}_{-0.11}$ & Mg~XI (blended) \\
 & & norm & $(2.44^{+1056}_{-0.41}) \times 10^{-6}$ & \\
\texttt{xszgauss} & Emission Line & $E_{\rm rest}$ & $1.85^{+0.02}_{-0.04}$ & Si~XIII (blended) \\
 & & norm & $(1.18^{+11.34}_{-0.21}) \times 10^{-6}$ & \\
\texttt{xszgauss} & Emission Line & $E_{\rm rest}$ & $2.32^{+0.04}_{-0.05}$ & S~XV (blended)\\
 & & norm & $(8.39^{+27.37}_{-2.13}) \times 10^{-7}$ & \\
\texttt{xszgauss} & Emission Line & $E_{\rm rest}$ & $2.88^{+0.10}_{-0.10}$ & S~XV \\
 & & norm & $(6.65^{+10.55}_{-2.22}) \times 10^{-7}$ & \\
\texttt{xszgauss} & Emission Line & $E_{\rm rest}$ & $7.14^{+0.06}_{-0.06}$ & Fe~XXVI \\
 & & norm & $(1.59^{+0.64}_{-0.58}) \times 10^{-6}$ & \\
\texttt{myts} & Scattered AGN Continuum (torus) & $N_{\rm H}$ & $2.53^{+0.02}_{-0.05}\times10^{23}$ & \\
 & & norm & $4.98^{+0.80}_{-0.28}\times10^{-3}$ & \\
\texttt{mytl} & Fluorescence Lines (torus) & norm & $1.12^{+0.37}_{-0.46}\times10^{-3}$ & \\
\bottomrule
\end{tabular}
\tablecomments{``norm'' refers to the normalization of the corresponding model component.}
\end{table*}

\begin{description}
    \item[~~FIRST~J120041]  
    The spectrum of this source is dominated by soft X-ray emission below 2~keV. An absorbed power law gives a good representation of the continuum, while additional Gaussian components are required at $E_{\rm rest} = 0.75^{+0.15}_{-0.54}$~keV (O~VIII, likely blended with O~VII), $E_{\rm rest} = 1.30^{+0.04}_{-0.05}$~keV (Mg~XI, likely blended with Mg~XII/Fe~L) and $E_{\rm rest} = 1.84^{+0.04}_{-0.04}$~keV (Si~XIII, possibly blended with Si~XIV). The best-fit line-of-sight absorption is $N_{\rm H} = 2.69^{+0.94}_{-0.96}\times10^{22}$~cm$^{-2}$ (best-fit model Cstat/d.o.f. = 2.92).  

    \item[~~2MASX~J130038]  
    The spectrum shows a pronounced hump between 3-6~keV, along with distinct emission features at both lower and higher energies. The best-fit model (Cstat/d.o.f. = 0.76) requires a \texttt{myts} scattered continuum, its associated \texttt{mytl} fluorescent line component, and an additional Gaussian at $E_{\rm rest} = 7.14^{+0.06}_{-0.06}$~keV. Below 3~keV, several Gaussian lines are detected at $E_{\rm rest} = 1.30^{+0.02}_{-0.11}$~keV (Mg~XI, possibly blended with Mg~XII), $1.85^{+0.02}_{-0.04}$~keV (Si~XIII/Si~XIV), $2.32^{+0.04}_{-0.05}$~keV (S~XV/S~XVI), and $2.88^{+0.10}_{-0.10}$~keV (S~XV). The 7.14~keV feature is discussed further in Section~\ref{sec:ufo_emission} as a candidate blueshifted Fe~XXVI~Ly$\alpha$ line. The best-fit line-of-sight absorption is $N_{\rm H} < 3.50\times10^{22}$~cm$^{-2}$.
\end{description}

\subsection{Physically Motivated Models}
\label{sec:physical}

To investigate the physical origin of the X-ray emission, we next fit models composed of photoionized and, when necessary, thermal plasma components. The photoionized emission is modeled using \texttt{Cloudy} \texttt{C.23} \citep{ferland_cloudy_1998, chatzikos_2023_2023}, while thermal emission is tested with \texttt{Apec} \citep{foster_updated_2012}. All fits adopt the phenomenological continuum from Section~\ref{sec:pheno}, with the intrinsic absorbing column allowed to vary above the Galactic value.

The \texttt{Cloudy} grids assume a broken power-law spectral energy distribution (SED) of the form $F_{\nu}=L\nu^{-\alpha}$, with $\alpha = 1.0$ below 13.6~eV, $\alpha=1.4$ between 13.6~eV and 0.5~keV, $\alpha=1.0$ from 0.5-10~keV, and $\alpha=0.5$ from 10-100~keV, adopting 1.4$\times$ solar abundances \citep[][see also Section~\ref{sec:cloudy_models}]{trindade_falcao_hubble_2021-1}. The final grids span $\log U = [-2,4]$ in steps of 0.25~dex and $\log N_{\rm H}=[19,23.5]$ in 0.1~dex steps. The spectra are exported as \texttt{XSPEC}-style additive tables \citep{porter_cloudyxspec_2006} and fitted using \texttt{Sherpa}.

A physical requirement of the photoionization models is that the emitting slab must be geometrically thin at the distance of the gas. Specifically, the slab thickness $\Delta r = N_{\rm H}/n_{\rm H}$ should satisfy $\Delta r \ll r$ \citep[e.g.,][]{kraemer_physical_2015}. This ensures that the cloud does not exceed either the amount of material available at that radius or the maximum emitting surface implied by the morphology. In practice, this condition introduces a coupled constraint between the ionization parameter ($\log U$) and the allowed range of column densities ($\log N_{\rm H}$): for a given $\log U$, only certain $\log N_{\rm H}$ are physically plausible.

Because $\log U$ varies during the fitting process, the upper limit on the column density, $\log N_{\rm H,\,max}$, cannot be imposed as a single fixed boundary. Instead, we implement a dynamic constraint in \texttt{Sherpa} that updates the maximum allowed column density at each iteration of the fit. The constraint has the general form
\begin{equation}
    \log N_{\rm H,\,max}(U) = a - b\,\log U ,
\label{eq:lohnh_max}    
\end{equation}

\noindent motivated by the $\Delta r \ll r$ condition and calibrated to the geometry of each region. In \texttt{Sherpa}, this relation is enforced by a helper function that updates the parameter bounds for the table model. By updating the allowed $\log N_{\rm H,\,max}$ range dynamically, the fit explores only the region of parameter space that corresponds to physically realizable clouds at the measured radial distances. This produces stable and self-consistent fits while preventing artificial degeneracies between $\log U$ and $\log N_{\rm H}$. 

The physically motivated models were constructed using the same iterative strategy and statistical criteria adopted for the phenomenological fits in Section~\ref{sec:pheno}. Starting from the minimal photoionized model required to reproduce the dominant soft X-ray emission, additional components (either \texttt{Cloudy} slabs or a thermal \texttt{Apec} plasma) were introduced one at a time and retained only if their inclusion resulted in a statistically meaningful improvement of the fit and a clear reduction of coherent residuals, as described in the following.

For each source, we explored models including up to three photoionized components, as well as combinations of photoionized and thermal plasma components. As in the phenomenological analysis, model complexity was increased only when justified by the data, adopting the same $\Delta$Cstat $\gtrsim 9$ criterion described in Section~\ref{sec:pheno} as a conservative empirical threshold for model inclusion. Components that did not meet this criterion, or whose parameters became unconstrained or degenerate with existing components, were not retained.

In FIRST~J120041, the data require two distinct photoionized components to reproduce the observed spectrum, whereas the addition of further photoionized or thermal plasma components does not yield a statistically significant improvement nor remove structured residuals. In contrast, the soft X-ray emission of 2MASX~J130038 is adequately described by a single photoionized component. In both sources, no thermal plasma component is required by the data, indicating that photoionization dominates the soft X-ray emission. The adopted models therefore provide a sufficient and physically self-consistent description of both spectra.

\begin{figure*}
  \centering
  \begin{minipage}[t]{0.49\textwidth}
    \centering
    \includegraphics[width=\linewidth]{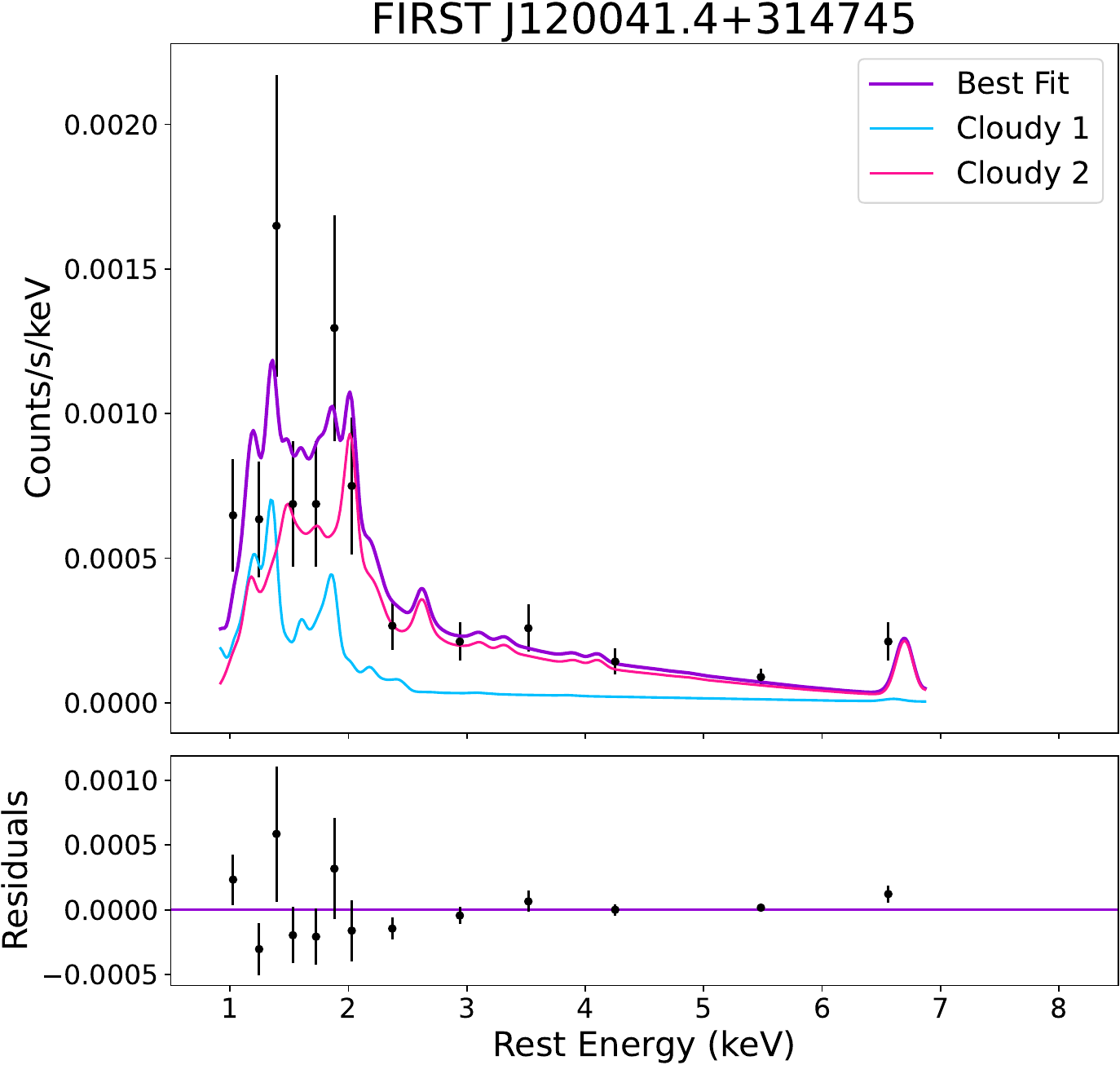}
  \end{minipage}%
  \hfill
  \begin{minipage}[t]{0.49\textwidth}
    \centering
    \includegraphics[width=\linewidth]{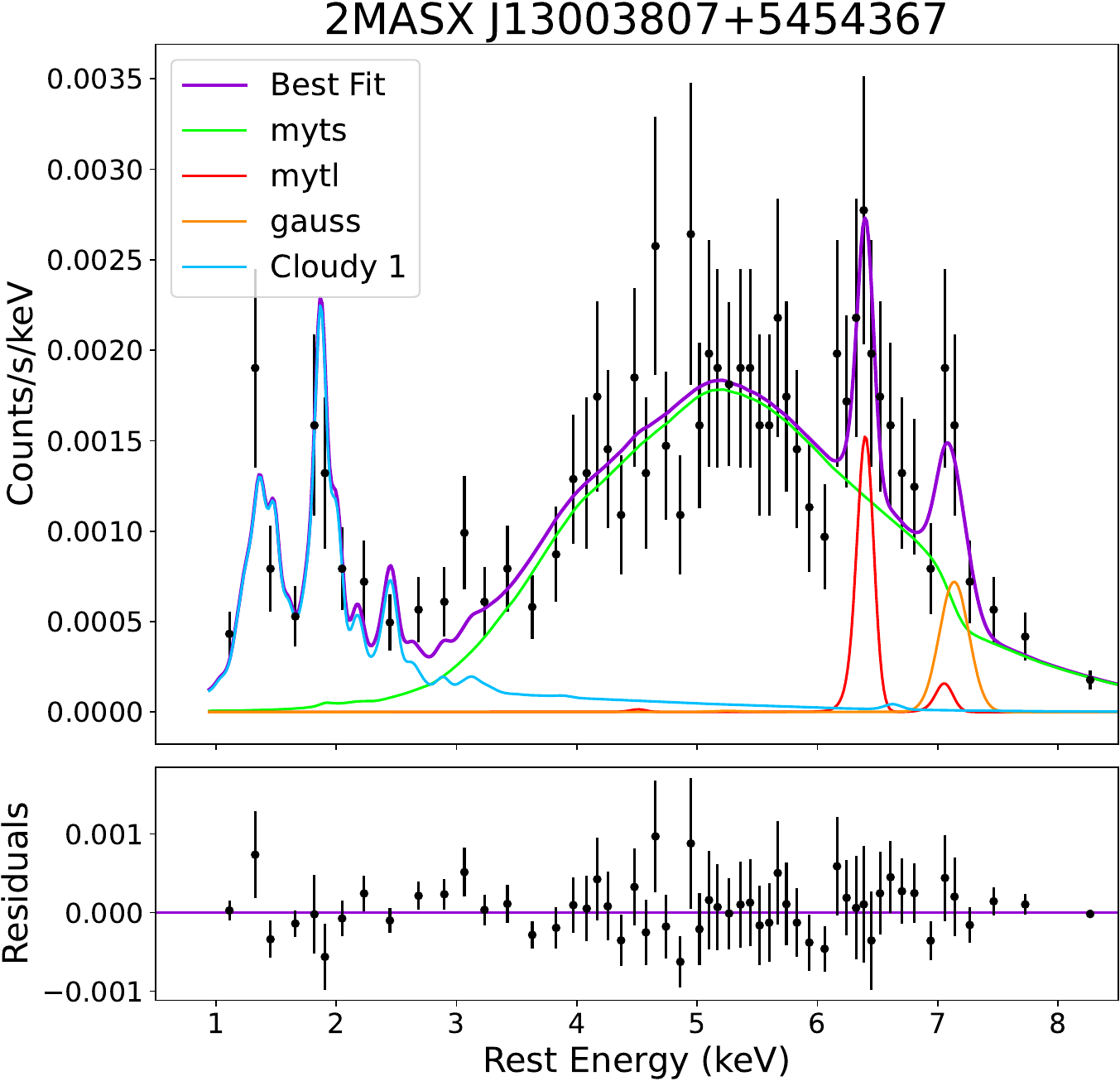}
  \end{minipage}
\caption{Rest-frame 0.3-8~keV spectra of FIRST~J120041 (left) and 2MASX~J130038 (right) fitted with photoionized components from \texttt{Cloudy}. Top panels show the model fits; bottom panels show residuals.}
 \label{fig:photo_thermal_composite} 
 \end{figure*}

The resulting best-fit physical models are presented in Figure~\ref{fig:photo_thermal_composite}, and their parameter values in Table~\ref{tab:spec_results}. Here and throughout the paper, parameters labeled as “n.c.” indicate values that are formally unconstrained by the data within the physically allowed parameter space.

\begin{description}
    \item[~~FIRST~J120041]  
    The spectrum is best described (Cstat/d.o.f. = 1.73) by one medium-ionization ($\log{U} = 0.49^{+0.40}_{-0.69}$) and medium-column ($\log{N_{\rm H}} = 21.92^{\rm n.c.}_{-0.14}$~cm$^{-2}$), and one high-ionization ($\log{U} = 2.5^{+0.82}_{-0.48}$) and low-column ($\log{N_{\rm H}} = 19.92^{+0.44}_{\rm n.c.}$~cm$^{-2}$) photoionized components. No additional continuum is required to reproduce the data.  

    \item[~~2MASX~J130038]  
    The best-fit model (Cstat/d.o.f. = 0.76) includes a scattered continuum (\texttt{myts}), the Gaussian line at 7.14~keV, and one medium-ionization ($\log{U} = 0.99^{+0.30}_{-0.39}$), medium-column ($\log{N_{\rm H}} = 21.09^{+0.12}_{\rm n.c.}$~cm$^{-2}$) photoionized component dominating the soft band.  
\end{description}

\begin{table}
\centering
\caption{Best-fit parameters from physically motivated models. Fits include one or more \texttt{Cloudy} components plus intrinsic absorption and the best-fit phenomenological continuum. Quoted errors are 1$\sigma$.}
\label{tab:spec_results}
\begin{tabular}{llc}
\toprule
Component & Parameter & Value \\
\midrule
\multicolumn{3}{c}{\textbf{FIRST~J120041.4+314745}} \\
\midrule
\multicolumn{3}{c}{2 photoionized components (Cstat/d.o.f. = 1.73)} \\
\hline
\texttt{xstbabs} & $N_{\rm H}$ & $(2.14^{+2.30}_{\rm n.c.})\times10^{21}$ \\
\texttt{CLOUDY 1} & $\log{U}$ & $0.49^{+0.40}_{-0.69}$ \\
 & $\log{N_{\rm H}}$ & 21.92$^{\rm n.c.}_{-0.14}$ \\
 & norm & $(3.89^{+9.27}_{-3.76})\times10^{-18}$ \\
\texttt{CLOUDY 2} & $\log{U}$ & $2.50^{+0.82}_{-0.48}$ \\
 & $\log{N_{\rm H}}$ & 19.92$^{+0.44}_{\rm n.c.}$ \\
 & norm & $(1.66^{+0.45}_{-0.65})\times10^{-17}$ \\

\midrule
\multicolumn{3}{c}{\textbf{2MASX~J13003807+5454367}} \\
\midrule
\multicolumn{3}{c}{1 photoionized component (Cstat/d.o.f. = 0.76)} \\
\hline
\texttt{xstbabs} & $N_{\rm H}$ & $(1.04^{+0.21}_{-0.20})\times10^{22}$ \\
\texttt{CLOUDY 1} & $\log{U}$ & $0.99^{+0.30}_{-0.39}$ \\
 & $\log{N_{\rm H}}$ & $21.09^{+0.12}_{\rm n.c.}$ \\
 & norm & $(2.68^{+0.65}_{-0.54})\times10^{-17}$ \\
\bottomrule
\end{tabular}
\tablecomments{${\rm n.c.}$ indicates that the parameter is not constrained from the fit.}
\end{table}

\section{Morphological Properties of the X-ray Emission}
\label{sec:morphology}

\subsection{PSF Modeling}
\label{sec:psf_model}

Accurate modeling of the Chandra point spread function (PSF) is essential for distinguishing real extended emission from unresolved nuclear components. For each observation, we simulate the PSF using the Chandra Ray Tracer (\texttt{ChaRT}\footnote{\url{https://cxc.cfa.harvard.edu/ciao/PSFs/chart2/index.html}}) and projected the rays onto the detector plane with \texttt{MARX}\footnote{\url{https://space.mit.edu/cxc/marx/}}. We generate ensembles of 1,000 PSF realizations centered at the source position and merge them within the same energy bands adopted for the imaging analysis (Section~\ref{sec:imaging}).

Following the empirical method of \citet{fabbiano_revisiting_2020}, each simulated PSF is normalized to the observed counts within an annulus of 0.2$''$-0.5$''$ radius. This normalization assumes that the observed emission comprises a central point-like component plus any additional extended structure. By testing multiple normalization radii we ensure robust scaling and minimize the risk of over-subtracting flux from the PSF wings.

\subsection{Broad- and Narrow-Band Imaging}
\label{sec:imaging}

We construct PSF-subtracted images in both broad and narrow energy bands. The narrow bands are selected to isolate line-dominated regions: 0.8-1.2~keV (Ne~IX-X; “Neon band”), 1.2-1.6~keV (Mg~XI-XII; “Magnesium band”), and 1.6-2.0~keV (Si~XIII-XIV; “Silicon band”). The broad bands are defined as 0.3-3~keV (“soft band”), 3-8~keV (“hard band”), and 0.3-8~keV (“full band”). For 2MASX~J130038, where the spectrum shows additional complexity at energies above 3~keV, we also produce images covering the 3-6~keV and Fe~K (6-8~keV) bands.

\begin{figure*}
  \centering
  \includegraphics[width=18cm]{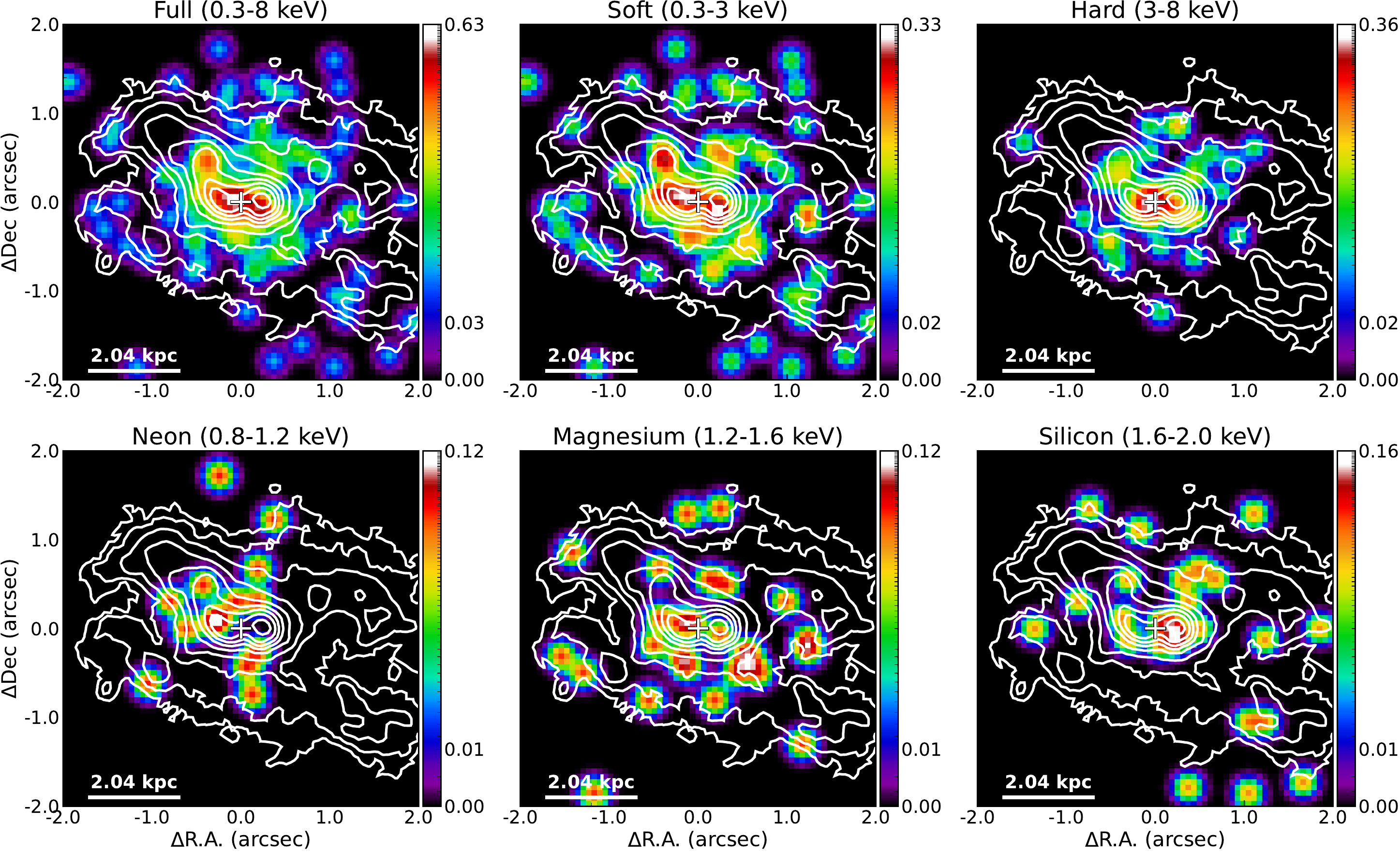}
  \caption{Chandra subpixel (1/8th of the native ACIS pixel) images of FIRST~J120041 in the indicated energy bands. North is up and east is left. PSF contributions have been subtracted; images are smoothed with a 3-pixel Gaussian and displayed using an asinh stretch. The color scale shows the X-ray counts per pixel. The hard-band centroid is marked with a white cross. White contours trace the [O~III] $\lambda5007$ emission from \emph{HST} imaging \citep{fischer_hubble_2018}, starting at $3\sigma$ above the background and increasing in powers of 2$\times$3$\sigma$.}
  \label{fig:hst_FIRST} 
\end{figure*}

\begin{figure*}
  \centering
  \includegraphics[width=18cm]{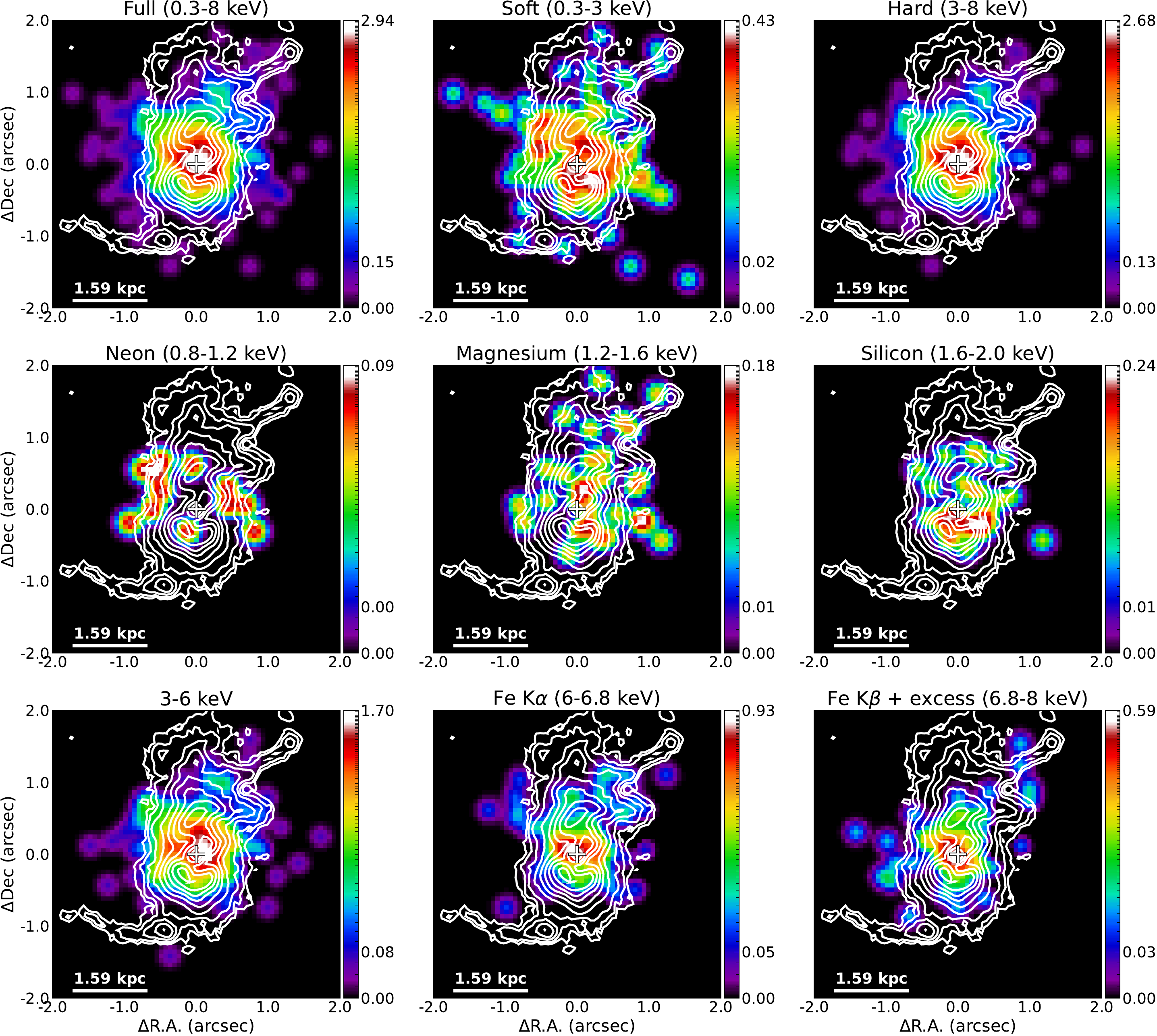}
\caption{Same as Figure~\ref{fig:hst_FIRST}, but for 2MASX~J130038.}
 \label{fig:hst_2MASX} 
 \end{figure*}

Figures~\ref{fig:hst_FIRST} and \ref{fig:hst_2MASX} show the resulting PSF-subtracted images, smoothed with a Gaussian kernel ($\sigma=1.5$) and displayed with asinh scaling. White contours trace the [O~III] $\lambda5007$ emission from archival \emph{HST} narrow-band imaging \citep{fischer_hubble_2018}. The contours start at $3\sigma$ above the background and increase in powers of two (i.e., $3\sigma \times 2^n$), such that the outermost contour marks the lowest significant [O~III] emission and successive inner contours correspond to progressively higher emission-line surface brightness. This allows for a direct, quantitative comparison between the X-ray and optical morphologies.

\begin{figure*}
  \centering
  \includegraphics[width=18cm]{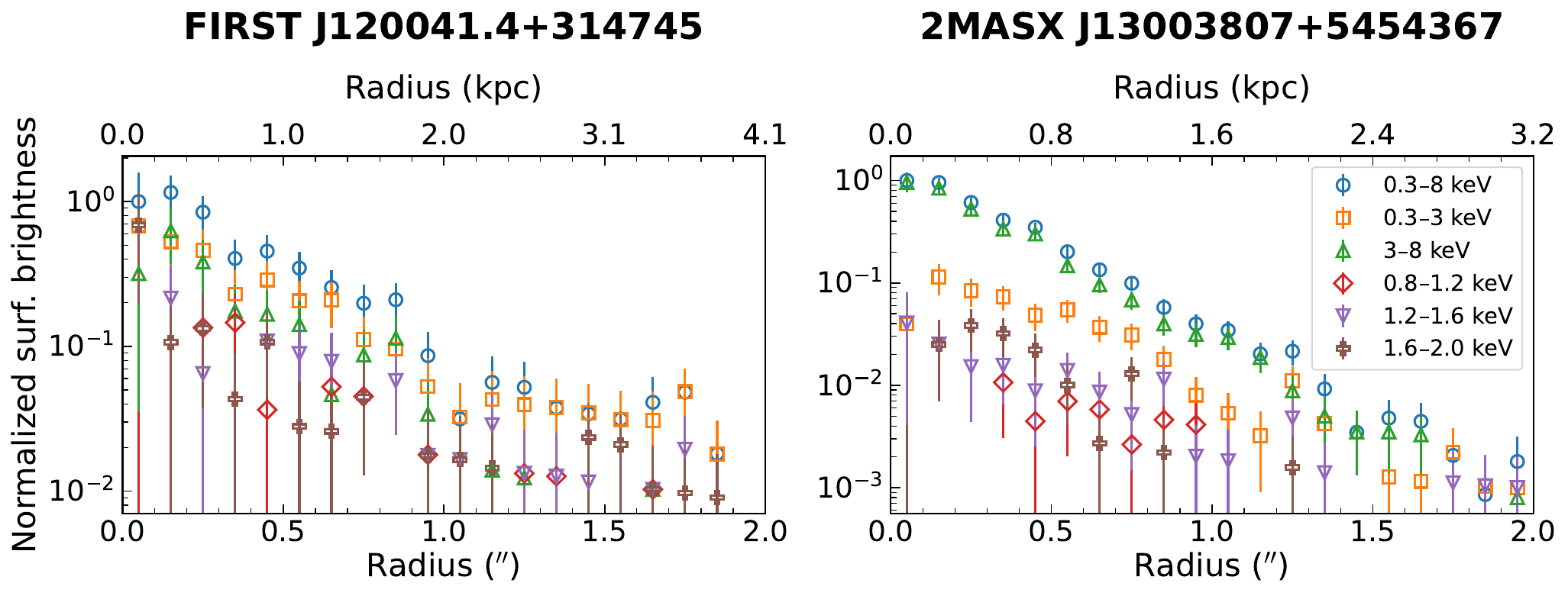}
\caption{Surface brightness radial profiles derived from PSF-subtracted narrow-band \textit{Chandra} images in the indicated energy bands. The profiles were extracted using 20 concentric annuli centered on the hard X-ray centroid and extending out to $r=2''$. They represent the total X-ray emission (continuum plus line contributions) within each bandpass and are not spectrally decomposed.}
 \label{fig:radial_prof} 
 \end{figure*}

In Figure~\ref{fig:radial_prof}, we show the X-ray surface brightness radial profiles for both quasars across several narrow energy bands. The profiles were extracted from the un-smoothed, PSF-subtracted Chandra images using 20 concentric annuli centered on each AGN and spanning radii of $0 \le r \le 2''$. To allow for direct comparison among bands, all curves are normalized to the surface brightness of the innermost bin in the full (0.3–8~keV) band.

\begin{description}
    \item[~~FIRST~J120041]  
    The full-band (0.3-8~keV) emission extends to $r\sim2''$ (4.1~kpc). The soft X-rays dominate, showing clumpy structures in multiple directions and two distinct peaks are evident: one located $\sim$0.2$''$ (410~pc) east of the centroid, likely associated with the peaks seen in the Neon and Magnesium band images, and another $\sim$0.2$''$ (410~pc) southwest of the centroid, consistent with the peak observed in the Silicon band. The soft-band morphology is spatially consistent with the [O~III] emission, with the brightest X-ray knots preferentially located within the higher-significance [O~III] contours. The hard-band emission is less extended ($r\sim1.4''$, 2.9~kpc) but shows similar correspondence with the inner [O~III] structures. In the narrow bands, the Neon band has a low signal-to-noise ratio (15 counts within 3.5$''$) but still shows extended emission peaking $\sim$0.35$''$ (715~pc) northeast of the nucleus. The Magnesium band (28 counts) shows a clumpy feature $\sim$0.6$''$ (1.2~kpc) southwest of the nucleus, while the Silicon band (25 counts) peaks west of the nucleus, coincident with the brightest [O~III] peak.  

    \item[~~2MASX~J130038]  
    The emission extends to $r\sim2''$ (3.2~kpc) but shows a simpler morphology and weaker correspondence with the [O~III] contours. The full-band image shows a bright, compact nucleus surrounded by fainter extended emission. The soft band reveals three distinct regions of enhanced emission: one located $\sim$0.6$''$ northeast of the centroid, another $\sim$0.2$''$ to the southeast, and the brightest region positioned $\sim$0.3$''$ southwest of the centroid. The remaining soft X-ray emission appears as faint clumps extending in multiple directions. The hard band is comparably extended ($\sim2''$) but dominated by a central component. In the narrow bands, the Neon band (12 counts within 2.5$''$) is confined within $\sim1''$. The Magnesium band (32 counts) shows peaks both above and below the nucleus, with a faint clump $\sim$0.9$''$ (1.4~kpc) southwest. The Silicon band (28 counts) peaks $\sim$0.4$''$ (640~pc) southwest, consistent with the soft-band maximum. At higher energies, the 3-6~keV emission resembles the full-band morphology, while the Fe~K$\alpha$ band (6-6.8~keV) is extended to $\sim1''$ (1.6~kpc) along the east-west axis. Emission in the 6.8-8~keV band is also extended, oriented northeast-southwest and confined within the inner 0.5$''$ (800~pc).
\end{description}

In all panels, the color scale represents the X-ray counts per pixel, displayed using a sinh$^{-1}$ stretch to enhance low-surface-brightness emission.
 
\section{X-ray Gas Modeling}
\label{sec:mod_xray}

\subsection{Photoionization Modeling}
\label{sec:cloudy_models}

We construct photoionization grids using \texttt{Cloudy} version~\texttt{C.23} \citep{chatzikos_2023_2023}. The predicted line spectrum depends on the shape of the ionizing SED, the gas density ($n_{\rm H}$), column density ($N_{\rm H}$), distance from the ionizing source ($r$), and the adopted chemical abundances. The models are parameterized by the dimensionless ionization parameter

\begin{equation}
    U=\frac{Q}{4\pi r^{2}c~n_{\rm H}}
\label{eq:ion_param}
\end{equation}

\noindent where $c$ is the speed of light and the ionizing photon rate $Q$ is given by

\begin{equation}
    Q = \int_{13.6~\mathrm{eV}}^{\infty} \frac{L_\nu}{h\nu} d\nu
\label{eq:q}
\end{equation}

Following Section~\ref{sec:physical}, we adopt an SED of the form $F_{\nu}=L\nu^{-\alpha}$, with $\alpha = 1.0$ below 13.6~eV, $\alpha=1.4$ between 13.6~eV and 0.5~keV, $\alpha=1.0$ from 0.5–10~keV, and $\alpha=0.5$ from 10–100~keV, with a high-energy cutoff at 100~keV. The ionizing luminosities are taken from \citet{trindade_falcao_hubble_2021-1}, based on the \textit{NuSTAR} analysis of \citet{gandhi_nustar_2014}: $Q=1.7\times10^{55}$~photons~s$^{-1}$ for FIRST~J120041 and $Q=6.5\times10^{54}$~photons~s$^{-1}$ for 2MASX~J130038 (but see Section~\ref{sec:model_line_lum}). Chemical abundances are set to $1.4\times$ solar, motivated by photoionization studies of AGN NLRs, which generally indicate modestly supersolar metallicities. In particular, \citet{groves_dusty_2004} inferred nitrogen abundances of approximately twice solar in NLR gas; since nitrogen is a secondary element whose abundance scales as $(Z/Z_\odot)^2$ \citep{talbot_evolution_1973}, this corresponds to an overall metallicity of approximately 1.4$\times Z_\odot$, which we adopt here \citep[e.g.,][]{trindade_falcao_hubble_2021-1}.

Because the emitting regions lie at distances of tens of parsecs or more, we adopt an open-slab geometry, following previous studies of extended photoionized gas \citep[e.g.,][]{kraemer_physical_2015, revalski_quantifying_2018, kraemer_mass_2020, trindade_falcao_hubble_2021-1}.

\subsection{Modeled Line Luminosities}
\label{sec:model_line_lum}

To convert the observed line fluxes into gas masses, we require estimates of both the emitting volume and the corresponding hydrogen density. For a photoionized cloud at radius $r$, the density is

\begin{equation}
    n_{\rm H} = \frac{Q}{4\pi cr^{2}U},
\label{eq:gas_density}
\end{equation}

At a fixed distance from the nucleus, $r=R$, the only free parameters in Equation~\ref{eq:gas_density} are $U$ and $n_{\rm H}$. For each model in the \texttt{Cloudy} grid, we therefore select a range of $U$ values that reproduce the observed emission lines and solve for the physically consistent density. 

We adopt $R=0.5''$, the characteristic average radius of the extraction region, for both quasars. Although the X-ray emission is spatially resolved in surface brightness (Section~\ref{sec:morphology}; Figure~\ref{fig:radial_prof}), the radial profiles are based on total, band-integrated counts and do not isolate individual emission lines. Determining gas densities and masses requires spectrally decomposed line luminosities and constraints on the ionization parameter in each radial bin. The photon statistics within individual annuli are insufficient to perform such independent spectral modeling as a function of radius. We are therefore restricted to a single, spatially integrated spectral measurement and cannot robustly model multiple radial zones or resolve a density gradient. The chosen radius represents the region enclosing a substantial fraction of the soft X-ray emission and provides a physically reasonable average distance for the line-emitting gas. We note that adopting a larger characteristic radius would decrease the inferred gas density as $n_{\rm H}\propto r^{-2}$ at fixed ionization parameter (see Equation~\ref{eq:gas_density}). However, to reproduce the observed line luminosities, the emitting area must increase proportionally as $A\propto r^{2}$. As a result, the total gas mass, which scales as $M \propto N_{\rm H} \times A$ (see Equation~\ref{eq:mass}), is largely insensitive to the assumed radius. Consequently, even if the line-emitting gas is distributed over larger radii, the derived mass estimates remain robust within the uncertainties.

FIRST~J120041 shows emission from multiple ionization states, requiring two photoionized components, hereafter \texttt{LOW} and \texttt{HIGH}. For simplicity, we assume both components are co-spatial (i.e., $R=0.5''$). In contrast, the spectrum of 2MASX~J130038 is well described by a single photoionized component (\texttt{HIGH}), which we treat as the sole contributor to its observed line luminosities.

To identify the best-fitting model at each location, we compare the modeled and observed line ratios for all fractional combinations of \texttt{LOW} and \texttt{HIGH} components; an ideal match yields a ratio of unity. The ionizing luminosities $Q$ are uncertain by a factor of $\sim$4 \citep{trindade_falcao_hubble_2021-1}. For FIRST~J120041, the published value reproduces the observed luminosities, whereas for 2MASX~J130038 a value larger by a factor of $\sim$3.8 is required, still within the allowed range.

Physically, the models are constrained by two requirements: (1) the emitting slab thickness, $\Delta R = N_{\rm H}/n_{\rm H}$, cannot exceed the available gas column at that radius, and (2) the emitting area cannot exceed the maximum geometric surface subtended by the source (see also Section~\ref{sec:physical}).

\subsection{Mass of Ionized Gas}
\label{sec:mass_ion_gas}

We estimate the total mass of the photoionized X-ray-emitting gas by combining the observed line luminosities with \texttt{Cloudy} photoionization models (Section~\ref{sec:cloudy_models}). Following \citet{kraemer_physical_2015, kraemer_mass_2020}, the mass of gas at radius $r$ is

\begin{equation}
    M_{\rm x-ray} = N_{\rm H}\,\mu m_{p}\,
    \left( \frac{L_{\rm line,obs}}{F_{\rm line,mod}} \right)
\label{eq:mass}
\end{equation}

\noindent where $\mu = 1.4$ is the mean mass per proton, $m_{p}$ is the proton mass, $L_{\rm line,obs}$ is the observed line luminosity, and $F_{\rm line,mod}$ is the modeled line flux. The term in parentheses represents an effective emitting area, $A_{\rm x-ray}$; multiplying this area by $N_{\rm H}$ yields the total number of particles covering the source, which scales to mass via $\mu m_p$.

Observed fluxes are measured using \texttt{srcflux}\footnote{\url{https://cxc.cfa.harvard.edu/ciao/ahelp/srcflux.html}} and converted to luminosities using distances from \citet{fischer_hubble_2018}. Table~\ref{tab:line_luminosities} lists the observed and modeled line luminosities within $r=1''$ for each target.

For FIRST~J120041, the masses of the two ionization components are computed separately and summed, $M_{\rm \texttt{TOT}}=M_{\rm \texttt{LOW}}+M_{\rm \texttt{HIGH}}$. This is achieved by expressing the observed luminosity as

\begin{equation}
   L_{\rm line,obs} = (F_{\rm line,\texttt{LOW}}\times A_{\rm \texttt{LOW}}) + (F_{\rm line,\texttt{HIGH}}\times A_{\rm \texttt{HIGH}}),
\label{eq:luminosity_line} 
\end{equation}

\noindent where $F_{\rm line,\texttt{LOW}}$ and $F_{\rm line,\texttt{HIGH}}$ are the modeled fluxes of each component. 

We solve for the emitting areas by fitting the Neon, Magnesium, and Silicon bands simultaneously, using the \texttt{SciPy} \texttt{differential\_evolution}\footnote{\url{https://docs.scipy.org/doc/scipy/reference/generated/scipy.optimize.differential\_evolution.html}} algorithm with geometric constraints as bounds. The \texttt{differential\_evolution} method is a global, population-based optimization algorithm that explores the full parameter space without requiring initial guesses or assuming local convexity of the solution. This approach is well suited for our problem, where the emitting areas are subject to physical bounds and the parameter space may contain multiple local minima. Uncertainties are estimated by perturbing the observed luminosities within their measurement errors and repeatedly re-optimizing in a Monte Carlo framework.

For 2MASX~J130038, only one photoionized component is required (Section~\ref{sec:physical}), so only a single emitting area is fitted. The resulting areas and masses are summarized in Table~\ref{tab:x-ray_masses}.

\subsection{Comparison with Optical Ionized [O~III] Gas}
\label{sec:comparison_mass_optical}

We now compare the mass of the X-ray-emitting gas to that of the cooler ionized phase traced by [O~III]. For simplicity, we assume that each phase is uniformly distributed within its respective extraction region. Under this assumption, a direct comparison requires accounting for the fact that the two phases occupy different volumes. We therefore define a volume-normalized X-ray mass

\begin{equation}
    \bar{M}_{\rm x-ray} = \frac{V_{\rm [O~III]}}{V_{\rm x-ray}}\times M_{\rm x-ray}
\label{eq:norm_mass}    
\end{equation}

\noindent where $V_{\rm [O~III]}$ and $V_{\rm x-ray}$ are the emitting volumes of the [O~III] and X-ray phases, respectively. 

The emitting volumes are estimated assuming spherical geometries whose radii correspond to the extraction regions used for each phase. Specifically, $V_{\rm x-ray}$ is computed assuming a sphere of radius $r=1''$ for both targets, matching the X-ray spectral extraction region, while $V_{\rm [O~III]}$ is computed assuming spheres with radii equal to the extent of the [O~III] outflowing regions ($r=0.5''$ for FIRST~J120041 and $r=0.1''$ for 2MASX~J130038). The volume ratio in Equation~\ref{eq:norm_mass} therefore reflects the different physical scales probed by the X-ray and [O~III] measurements, rather than a pixel-by-pixel comparison within the common field of view, and allows a direct comparison of the relative contributions of the hot and warm ionized gas under a common geometric framework.

In reality, local gas density is expected to decrease with radius \citep[e.g.,][]{revalski_quantifying_2022}. Moreover, the [O~III] phase, having lower ionization and higher density than the X-ray-emitting gas, requires a correspondingly smaller volume to reproduce its observed luminosity. These effects imply that assuming uniform density and unity filling factor tends to overestimate the emitting volumes, especially for the X-ray gas. Consequently, the volume-normalized mass ratios derived below should be regarded as conservative \textit{upper limits}, particularly in systems where the X-ray-emitting gas extends over a substantially larger region than the [O~III] gas.

\begin{description}
    \item[~~FIRST~J120041]  
    Within $r=1''$ (2~kpc), the low- and high-ionization X-ray components contain $M_{\rm \texttt{LOW}} = 4.1^{+0.9}_{-0.8}\times10^{8}~M_{\odot}$ and $M_{\rm \texttt{HIGH}} = 2.8^{+6.0}_{-2.1}\times10^{5}~M_{\odot}$, respectively. The total X-ray mass is therefore dominated by the low-ionization phase: $M_{\rm xray} = 4.1^{+0.9}_{-0.8}\times10^{8}~M_{\odot}$. The [O~III] outflow contains $M_{\rm [O~III]} = 1.4\times10^{7}~M_{\odot}$ \citep{trindade_falcao_hubble_2021-1} within $r=0.5''$ (1~kpc). After accounting for the different emitting volumes defined by the respective extraction radii (Equation~\ref{eq:norm_mass}), the resulting volume-normalized mass ratio is $\bar{M}_{\rm x-ray}/M_{\rm [O~III]}\sim 4$.

    \item[~~2MASX~J130038]  
    Within $r=1''$ (1.6~kpc), the X-ray-emitting gas contains $M_{\rm x-ray} = 1.8^{+0.2}_{\rm n.c.}\times10^{8}~M_{\odot}$. The [O~III] outflow contains $M_{\rm [O~III]} = 1.1\times10^{4}~M_{\odot}$ within $r=0.1''$ (0.16~kpc). After correcting for the different emitting volumes, the resulting volume-normalized mass contrast is $\bar{M}_{\rm x-ray}/M_{\rm [O~III]}\sim 16$.
\end{description}

For comparison, nearby Seyfert galaxies host significantly smaller X-ray gas reservoirs. NGC~1068 contains $M_{\rm x-ray}\approx5.6\times10^{5}~M_{\odot}$ within the inner $r\sim100$~pc, combining low- and high-ionization components \citep{kraemer_physical_2015}, while NGC~4151 has $M_{\rm x-ray}\approx5.4\times10^{5}~M_{\odot}$, within $r\sim500$~pc \citep{kraemer_mass_2020}. The X-ray gas masses of FIRST~J120041 and 2MASX~J130038 exceed these values by factors of $\sim$300-800 (2.5-3~dex), though measured within volumes $\sim$30-8000 times larger.

\begin{table}
\centering
\caption{Observed and modeled X-ray line luminosities within the inner $r=$1$''$.}
\label{tab:line_luminosities}
\centering
\begin{tabular}{cccc}
\toprule
Band & Modeled & $L_{\rm line,obs}$ & $L_{\rm line,mod}$  \\
(keV) & Lines& ($10^{40}$~erg~s$^{-1}$) & ($10^{40}$~erg~s$^{-1}$) \\
\midrule
\multicolumn{4}{c}{\textbf{FIRST~J120041.4+314745}} \\
0.8-1.2 & Ne~X Ly$\alpha$ & $16^{+5.3}_{-4.2}$ & $8.7^{+2.0}_{-1.8}$   \\
1.2-1.6 & Mg~XI-XII & $5.7^{+73}_{-4.4}$ & $12.6^{+2.9}_{-2.6}$ \\
1.6-2.0 & Si~XIII-XIV & $3.7^{+1.1}_{-0.9}$ & $4.5^{+1.0}_{-0.9}$   \\
\midrule
\multicolumn{4}{c}{\textbf{2MASX~J13003807+5454367}} \\
0.8-1.2 & Ne~X Ly$\alpha$ & $11.2^{+3.9}_{-3.2}$ & $6.7^{+0.9}_{-0.8}$   \\
1.2-1.6 & Mg~XI-XII & $8.1^{+1.8}_{-1.5}$ & $7.5^{+0.9}_{-0.9}$ \\
1.6-2.0 & Si~XIII-XIV & $6.8^{+1.4}_{-1.2}$ & $7.8^{+1.0}_{-1.0}$   \\
\bottomrule
\end{tabular}
\tablecomments{Only emission lines with significant predicted contributions in the \texttt{Cloudy} models are included.}
\end{table}

\begin{table*}
\centering
\caption{Calculated effective emitting areas and mass of hot X-ray-emitting gas within
the inner $r=$1$''$.}
\label{tab:x-ray_masses}
\centering
\begin{tabular}{ccccc}
\toprule
$A_{\rm \texttt{LOW}}$ & $A_{\rm \texttt{HIGH}}$ & $M_{\rm \texttt{LOW}}$ & $M_{\rm \texttt{HIGH}}$ & $M_{\rm x-ray}$ \\
($10^{43}$~cm$^{2}$) & ($10^{43}$~cm$^{2}$) & ($10^{8}M_{\odot}$) & ($10^{5}M_{\odot}$) & ($10^{8}M_{\odot}$) \\
\midrule
\multicolumn{5}{c}{\textbf{FIRST~J120041.4+314745}} \\
4.5$^{+1.04}_{-0.93}$ & 0.31$^{+0.66}_{-0.23}$ & $4.1^{+0.95}_{-0.84}$ & $2.8^{+6.0}_{-2.1}\times10^{-3}$ & $4.1^{+0.95}_{-0.85}$\\
\midrule
\multicolumn{5}{c}{\textbf{2MASX~J13003807+5454367}} \\
-&4.7$^{+0.60}_{-0.59}$ & - & $1.8^{+0.24}_{\rm n.c.}$ & $1.8^{+0.24}_{\rm n.c.}$\\
\bottomrule
\end{tabular}
\tablecomments{$A_{\rm \texttt{LOW}}$, $A_{\rm \texttt{HIGH}}$, $M_{\rm \texttt{LOW}}$ and $M_{\rm \texttt{HIGH}}$ are the calculated effective emitting areas and masses for the low- and high-ionization model components.}
\end{table*}

\section{The Role of X-ray Gas}
\label{sec:role_xray}

\subsection{[O~III] Gas Kinematics}
\label{sec:kinematics}

To investigate whether the X-ray and optical phases are dynamically correlated, we compare the X-ray surface brightness with the kinematics of the ionized [O~III] gas \citep{fischer_hubble_2018}. In that study, the authors defined three kinematic classes based on the [O~III] velocity centroids relative to systemic and line widths, which we adopt here:

\begin{itemize}
    \item[$-$] \textbf{AGN-driven outflows:} [O~III] lines with centroid velocities $>300$~km~s$^{-1}$ or multiple kinematic components.
    \item[$-$] \textbf{Disturbed rotation:} Single-component [O~III] lines with FWHM~$>250$~km~s$^{-1}$ but centroid velocities $<300$~km~s$^{-1}$.
    \item[$-$] \textbf{Non-disturbed rotation:} [O~III] lines with centroid velocities $<300$~km~s$^{-1}$ and FWHM~$\lesssim250$~km~s$^{-1}$, consistent with ordered disk rotation \citep[see also][]{bellocchi_vltvimos_2013, ramos_almeida_infrared_2017}.
\end{itemize}

\begin{figure*}
  \centering
  \includegraphics[width=\textwidth]{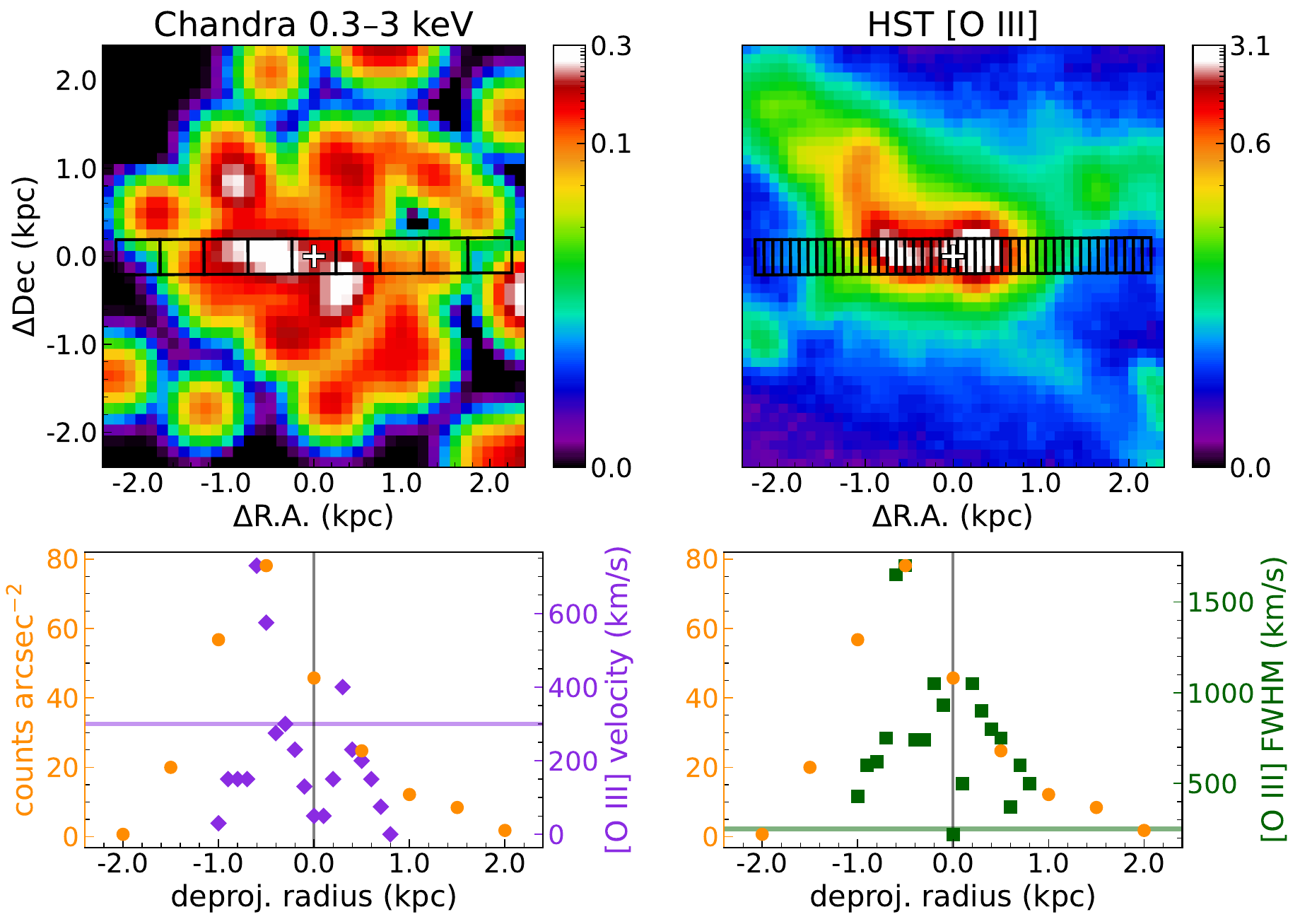}
  \caption{\textit{Top:} Chandra 0.3-3~keV image of FIRST~J120041 (left) and HST [O~III] image with STIS slit (right) shown in log scale. Color bars are in units of counts (left) and 10$^{-17}$~erg~s$^{-1}$~cm~$^{-2}$ (right). \textit{Bottom:} Radial soft X-ray surface brightness profile (deprojected, orange) compared with deprojected [O~III] velocities (purple, left) and FWHM line widths (green, right). Horizontal lines mark the threshold for outflowing gas (purple, $>$300~km~s$^{-1}$) and disturbed gas kinematics (green, $>$250~km~s$^{-1}$) \citep{fischer_hubble_2018}.}
  \label{fig:hst_FIRST_slit} 
\end{figure*}

\begin{figure*}
  \centering
  \includegraphics[width=\textwidth]{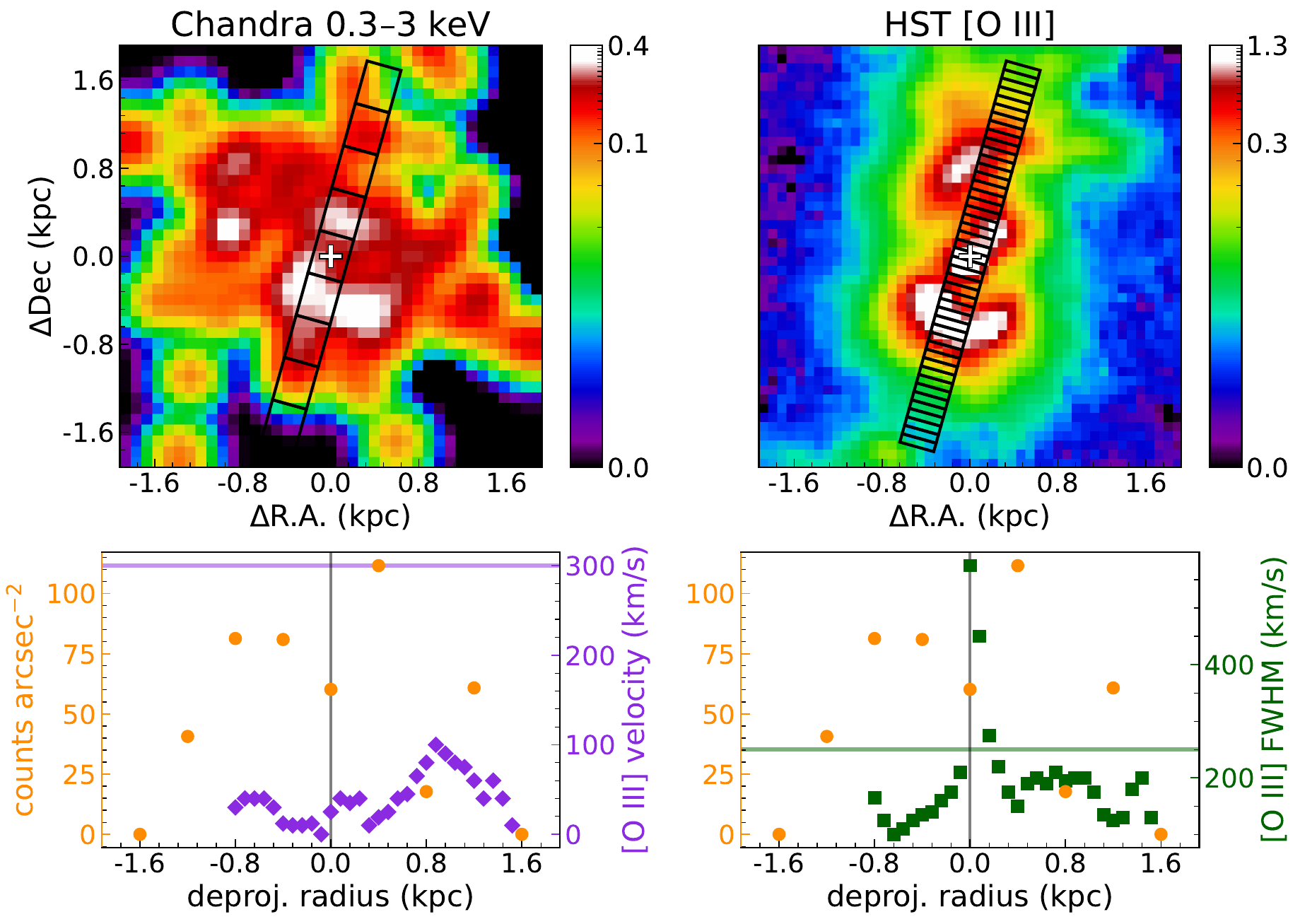}
  \caption{Same as Figure~\ref{fig:hst_FIRST_slit}, but for 2MASX~J130038.}
  \label{fig:hst_2MASX_slit} 
\end{figure*}

We use the \texttt{dmextract}\footnote{\url{https://cxc.cfa.harvard.edu/ciao/ahelp/dmextract.html}} task to measure the surface brightness of the 0.3-3~keV soft X-ray emission along rectangular extraction strips aligned with the HST/STIS slit positions from \citet{fischer_hubble_2018}. Each strip spans $2.25''$ in length and is divided into nine $0.25''$ bins, matching the angular resolution of Chandra. Figures~\ref{fig:hst_FIRST_slit} and \ref{fig:hst_2MASX_slit} show, respectively, the extraction geometry overlaid on the 0.3-3~keV Chandra images (top left) and the STIS slit over the HST [O~III] images (top right). The bottom panels compare the spatially resolved X-ray surface brightness profiles (orange) with the [O~III] velocity centroids (purple, bottom left) and [O~III] FWHM values (green, bottom right) for both quasars.

\begin{description}
    \item[~~FIRST~J120041]  
    The soft X-ray surface brightness profile confirms that the emission peaks off-nucleus, with a clear enhancement to the east of the nucleus ($r\sim500$~pc). Out to at least $1$~kpc in that direction, the X-ray surface brightness remains higher than at the nuclear position. Comparing this distribution to the [O~III] \textit{outflowing} kinematics (purple points above the reference line in the bottom-left panel), we find that regions of enhanced X-ray surface brightness ($>$40~counts~arcsec$^{-2}$) coincide with locations where the [O~III] gas shows the highest velocities ($\sim600$-750~km~s$^{-1}$) and the broadest line widths (up to $\sim$1700~km~s$^{-1}$). This spatial correspondence shows that the brightest, hot X-ray-emitting gas is co-spatial with the outflowing [O~III] component, suggesting a direct interaction between the two phases. While the [O~III] flux is higher in the right side (western) emission knot, the outflow velocities and FWHMs are greater on the left (eastern) side, exactly where the soft X-ray peak happens. This further supports the interpretation that the enhanced X-ray emission pinpoints regions of strong coupling between the hot ionized and the cooler [O~III] outflow detected by \citet{fischer_hubble_2018}.

    \item[~~2MASX~J130038]  
    The soft X-ray surface brightness profile also peaks off-nucleus, showing enhancements on both sides of the nucleus. Out to $\sim$1~kpc in either direction along the slit, the extended emission has higher surface brightness than the nuclear region. Although the per-bin X-ray surface brightness (orange points in Figure~\ref{fig:hst_2MASX_slit}) is higher than in FIRST~J120041, the [O~III] kinematics in this quasar (purple points, bottom-left panel) are consistent with ordered rotation, i.e., not-outflowing \citep{fischer_hubble_2018}. The centroid velocities remain below the threshold for outflowing gas (purple line), and the FWHM values are generally below the disturbed gas threshold (green line), except for a localized broadened component $\sim160$~pc northwest of the nucleus.  
    This broadened region coincides with the base of the linear [O~III] structure observed near the nucleus. In the X-ray image, an enhancement appears at the end of this structure (see also Figure~\ref{fig:hst_2MASX}) and another to the southeast, at the location of the prominent [O~III] arc. The combination of high FWHM ($>$250~km~s$^{-1}$) but low centroid velocity ($<$300~km~s$^{-1}$) at these position falls into the “disturbed gas” category of \citet{fischer_hubble_2018}, suggesting that small-scale turbulence or nascent outflow activity could be arising from the linear [O~III] feature. Despite its relatively high X-ray surface brightness, no [O~III] outflows are detected in this system, and Figure~\ref{fig:hst_2MASX_slit} shows no clear spatial correlation between the X-ray enhancements and the [O~III] velocity or line-width distributions.
\end{description}

\subsection{A Relativistic Hot Wind?}
\label{sec:ufo_emission}

The spectrum of 2MASX~J130038 shows a strong neutral Fe~K$\alpha$ line at 6.4~keV (EW = 228~eV) together with an additional emission feature at $E_{\rm rest}=7.14$~keV (Figures~\ref{fig:pheno}, \ref{fig:photo_thermal_composite}). While the \texttt{mytl} component successfully reproduces both the neutral Fe~K$\alpha$ line at $E_{\rm rest}=6.39$~keV and the associated Fe~K$\beta$ line at $E_{\rm rest}=7.06$~keV, residual emission remains at higher energies, motivating the inclusion of an additional Gaussian component.

When adding a Gaussian line with width $\sigma\geq0.1$~keV and free centroid energy and normalization to the best-fit broadband continuum model derived in Sections~\ref{sec:pheno} and \ref{sec:physical}, with the continuum parameters held fixed, the fit improves by $\Delta$Cstat $=10$ for two additional free parameters. Under the asymptotic $\chi^2$ approximation, this corresponds to a $p$-value of $\sim7\times10^{-3}$ (i.e., $\sim2.7\sigma$). We therefore regard this feature as marginal but potentially meaningful, and treat it with appropriate caution in the discussion below.

If interpreted as blueshifted Fe~XXVI~Ly$\alpha$ emission (rest energy 6.97~keV), the observed centroid implies an outflow velocity of $\sim7,000$-$8,000$~km~s$^{-1}$. Such velocities are comparable to those inferred for highly ionized nuclear winds in luminous AGNs, which are typically detected in absorption \citep[e.g.,][]{tombesi_discovery_2010, tombesi_wind_2015}. Given the modest statistical significance of the feature, we refrain from drawing strong conclusions and consider this identification suggestive rather than definitive.

Alternative identifications are less satisfactory. Fe~K$\beta$ fluorescence from neutral iron (rest energy 7.06~keV) would require a flux well above the expected $\sim$10-15\% of the Fe~K$\alpha$ line \citep{molendi_iron_2003, bianchi_soft_2006}. Moreover, the \texttt{mytl} component already reproduces the expected Fe~K$\beta$ emission, with a fitted flux $\sim$16\% of the Fe~K$\alpha$ intensity (Figures~\ref{fig:pheno}, \ref{fig:photo_thermal_composite}). Emission from Ni~K$\alpha$ (rest energy 7.47~keV) lies too far from the measured centroid to provide a satisfactory explanation, although this possibility cannot be entirely excluded given the current data quality.

We therefore note the presence of a possible high-energy emission feature consistent with Fe~XXVI~Ly$\alpha$ from a fast, highly ionized nuclear wind, while emphasizing that deeper hard X-ray observations are required to confirm its nature and physical origin. In contrast to the ultra-fast outflows commonly reported in quasars, which are detected in absorption \citep[e.g.,][]{tombesi_discovery_2010, tombesi_wind_2015}, this candidate outflow would be observed in emission, potentially implying a geometry and covering factor that allow substantial reprocessing of the nuclear radiation field. We discuss the implications of such a scenario in the context of quasar evolution in Section~\ref{sec:evolution}.

\section{An Evolutionary Path for Quasars}
\label{sec:evolution}

The QSO2s analyzed by \citet{fischer_hubble_2018} span a wide range of kinematic and morphological properties. Their sample includes twelve of the fifteen most luminous type~2 quasars in the \citet{reyes_space_2008} catalog with $z<0.12$. All targets have $\log L_{\mathrm{[O~III]}}\geq42.28$~erg~s$^{-1}$, placing them in the brightest $\sim25\%$ of the $z<0.3$ QSO2 population from \citet{reyes_space_2008}. The corresponding black hole masses lie in the range $10^{7.7}$–$10^{8.3}~M_\odot$ \citep[e.g.,][]{fischer_hubble_2018}. Most sources are radio-quiet, with a minority classified as intermediately radio-loud, indicating that their observed outflows are unlikely to be jet-driven. Both quasars analyzed in this work, FIRST~J120041 and 2MASX~J130038, are classified as radio-quiet objects, and therefore belong to the radio-quiet majority of the sample.

All twelve objects host [O~III]-emitting NLRs that extend beyond 1~kpc, yet only two show outflows that propagate on similarly large scales. \citet{fischer_hubble_2018} argued that the distribution of circumnuclear gas near the nucleus may regulate how far ionizing radiation and winds can penetrate into the NLR: in morphologically compact systems, dense inner material can absorb a significant fraction of the AGN ionizing flux, confining outflows to smaller radii. Building on this picture, \citet{trindade_falcao_hubble_2021} suggested that inner X-ray winds may progressively entrain circumnuclear gas, an idea that has also been explored in theoretical models of AGN feedback \citep[e.g.,][]{costa_feedback_2014} and supported by observational studies tracing multi-phase outflows through X-ray absorption features \citep[e.g.,][]{serafinelli_multiphase_2019, xu_systematic_2024}. In this framework, continued entrainment may eventually enable a transition into a large-scale blowout phase \citep{trindade_falcao_hubble_2021}.

In this work, we propose that these quasars evolve along a \emph{continuous} sequence regulated by the progressive clearing of circumnuclear material by the AGN. In the earliest, most obscured phase, both ionizing radiation and winds are confined to the nuclear region. As the surrounding medium becomes more porous, low-density channels begin to form, allowing ionizing radiation to reach larger radii and enabling high-ionization winds to emerge from the inner few hundred parsecs. Once the winds fully break through, they can drive large-scale outflows capable of reshaping the host ISM.

\begin{figure*}
  \centering
  \includegraphics[width=\textwidth]{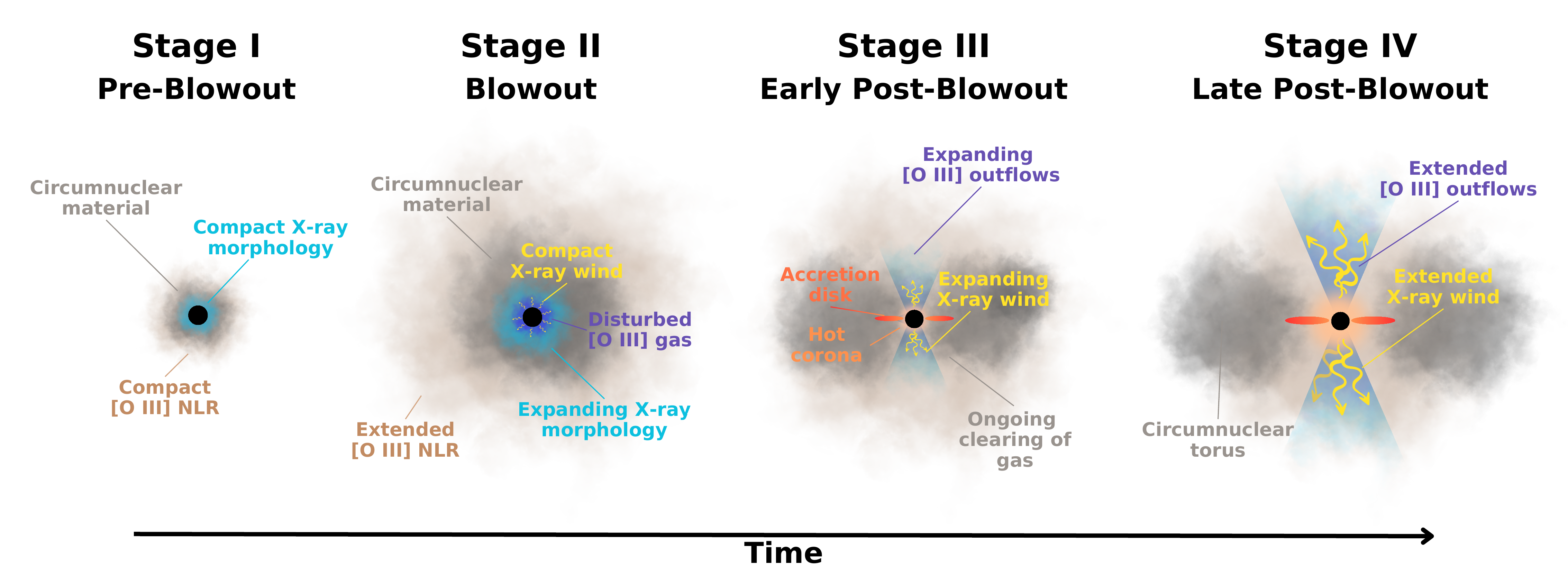}
 \caption{Schematic illustration of the proposed evolutionary sequence for quasars. The four labeled phases - (\textbf{I}) Pre-Blowout, (\textbf{II}) Blowout, (\textbf{III}) Early Post-Blowout, and (\textbf{IV}) Late Post-Blowout - represent illustrative steps along a continuous process of circumnuclear clearing. Each panel highlights the characteristic [O~III] and X-ray emission associated with increasing coupling between the hot X-ray and cooler ionized [O~III] gas phases as AGN winds progressively disperse the surrounding material.}
  \label{fig:comparison} 
\end{figure*}

The differing propagation timescales of ionizing radiation and mechanical feedback naturally account for systems with extended ionized [O~III] emission but no large-scale winds. A wind traveling at $v\sim1000$~km~s$^{-1}$ requires $\sim10^{6}$~yr to cross 1~kpc, whereas the light-crossing time to the same distance is only $\sim3\times10^{3}$~yr. Consequently, photoionization can illuminate kiloparsec-scale [O~III] structures long before mechanical winds have propagated to comparable radii. This evolutionary framework, summarized schematically in Figure~\ref{fig:comparison}, links the diversity of morphologies and kinematics observed by \citet{fischer_hubble_2018} to different stages along a continuum sequence of circumnuclear clearing and feedback coupling.

Although this framework is developed using a low-redshift QSO2 sample, it is not intended to be tied to fixed luminosities or physical scales. Instead, the proposed stages describe relative evolutionary stages regulated by obscuration and feedback coupling. At higher redshifts and luminosities, such as in Hot Dust-Obscured Galaxies (Hot DOGs; see, e.g., \citealt{eisenhardt_first_2012, assef_half_2015}), the same sequence may operate at larger energies and on shorter timescales. In this context, Hot DOGs can be interpreted as extreme analogs of the late pre-blowout and blowout phases, in which systems remain heavily obscured while already exhibiting powerful ionized outflows traced by [O~III], even as direct X-ray signatures of the inner wind are attenuated by large column densities \citep[e.g.,][]{costa_feedback_2014, serafinelli_multiphase_2019, xu_systematic_2024, villani_deep_2026}. Objects that fully reach a late post-blowout phase would likely no longer satisfy Hot DOG selection criteria.

Our Chandra observations support this proposed scenario (see Figure~\ref{fig:comparison}). The X-ray and [O~III] properties of 2MASX~J130038 and FIRST~J120041 suggest that they occupy different locations along the early portion of this evolutionary sequence. In 2MASX~J130038, the X-ray emission is extended but only weakly correlated with the [O~III] morphology, while the optical kinematics remain dominated by ordered rotation with localized turbulence near the nucleus. Combined with the tentative detection of a blueshifted Fe~XXVI line at $v\sim7600$~km~s$^{-1}$ (Section~\ref{sec:ufo_emission}), these features point to a system in the onset of a blowout stage (Stage~II; see Figure~\ref{fig:comparison}), in which a nascent hot wind is beginning to leak out of the nucleus and interact with the inner circumnuclear medium. 

In contrast, FIRST~J120041 shows clumpy, extended X-ray emission that spatially correlates with the [O~III] bicone. Peaks in X-ray surface brightness coincide with regions of the highest [O~III] velocities and line widths, pointing to a morphological and kinematic correspondence between the hot and cooler ionized phases. This correspondence suggests that the X-ray wind has begun to couple dynamically to the [O~III]-emitting gas, consistent with a relatively more developed stage of the sequence (Stage~III; see Figure~\ref{fig:comparison}).

Taken together, these quasars may trace the transition from a kinematically compact, blowout configuration (Stage~II) to the onset of a more extended, early post–blowout state (Stage~III). We emphasize that the characteristic spatial scales quoted below (e.g., $R_{\rm [O~III]}<3$~kpc for Stage~I or $R_{\rm [O~III]}\sim4$-6~kpc for later stages) refer specifically to the objects in the \citet{fischer_hubble_2018} sample discussed here. These values are intended as empirical descriptors rather than universal thresholds, and are expected to depend on intrinsic AGN properties such as black hole mass and luminosity. For clarity, in Figure~\ref{fig:comparison} we illustrate this sequence through four representative “stages,” while Table~\ref{tab:evol_sequence} summarizes how the twelve QSO2s studied by \citet{fischer_hubble_2018} may populate these stages. These are labeled as follows:

\begin{description}
\item[~~Stage~I: Pre-Blowout] This initial phase represents heavily enshrouded systems in which both the X-ray and [O~III] emission are morphologically compact and centrally concentrated. Ionizing radiation and winds remain trapped by dense circumnuclear gas, preventing large-scale outflows to eventually form. In the \citet{fischer_hubble_2018} sample, examples may include 2MASX~J11001238, 2MASX~J14054117, and 2MASX~J17135038, which all show compact [O~III] morphologies ($R_{\rm [O~III]}<3$~kpc; \citealt{fischer_hubble_2018}) but broad nuclear line widths (FWHM$\gtrsim$1000~km~s$^{-1}$; \citealt{fischer_hubble_2018}) indicative of turbulent but spatially confined gas in the inner region.

    \item[~~Stage~II: Blowout] Systems where the AGN ionizing radiation begins to escape through low-density channels, ionizing more extended structures but still lacking large-scale outflows. 2MASX~J130038 exemplifies this behavior: its [O~III] emission extends to $R_{\rm [O~III]}=4.7$~kpc, yet the gas is largely rotational, with only localized disturbances within $\sim$160~pc of the nucleus \citep{fischer_hubble_2018}. The X-rays are morphologically centrally peaked and weakly correlated with [O~III] emission (Figure~\ref{fig:hst_2MASX}), implying that the hot X-ray and cooler [O~III] phases remain decoupled. The tentative identification of Fe~XXVI emission at 7.14~keV ($v\sim7600$~km~s$^{-1}$; Section~\ref{sec:ufo_emission}) could point to the onset of a nascent hot wind beginning to interact with the inner circumnuclear gas, but is not required for the Stage~II classification. The evolutionary interpretation is instead driven primarily by the spatial decoupling between the X-ray and [O~III] emission and by the lack of large-scale outflow kinematics. Similar Stage~II behavior is seen in other systems such as 2MASX~J07594101, 2MASX~J080252, SDSS~J115245, B2~1435, Mrk~477, and 2MASX~J16531506, where [O~III] ionized emission is extended ($R_{\rm [O~III]}\gtrsim$3~kpc; \citealt{fischer_hubble_2018}) but the kinematics remain dominated by broad line widths and low velocity centroids (FWHM~$\gtrsim$600~km~s$^{-1}$, $v<$600~km~s$^{-1}$; \citealt{fischer_hubble_2018}). The large X-ray gas reservoir found in 2MASX~J130038 likely reflects the accumulation of material during the first evolutionary phases, when dense circumnuclear gas remained confined close to the AGN. As this gas lies nearer to the ionizing source, it reaches higher ionization states, naturally producing a more massive X-ray-emitting component at the onset of blowout.

    \item[~~Stage~III: Early Post-Blowout] As the wind begins to break through the circumnuclear material, the X-ray outflow expands, producing outflowing [O~III] kinematics and the spatially correlated X-ray and optical emission we observe. FIRST~J120041 shows bright [O~III] emission extending to $R_{\rm [O~III]}=6.1$~kpc, with multi-component, high-velocity profiles ($v\sim600$-750~km~s$^{-1}$) and broad line widths (FWHM~$\sim$1700~km~s$^{-1}$; \citealt{fischer_hubble_2018}). The soft X-ray emission is clumpy and structurally complex, showing a strong morphological correspondence with the [O~III] gas (Figure~\ref{fig:hst_FIRST}). Peaks in the soft X-ray brightness are co-spatial with regions of outflowing [O~III] gas (Figure~\ref{fig:hst_FIRST_slit}), suggesting that the X-ray wind is now entraining and driving the optical phase. These properties are consistent with a system in which a blowout has recently occurred, marking the transition from a confined to an expanding feedback regime.

    \item[~~Stage~IV: Late Post-Blowout] The nuclear X-ray winds have now cleared most of the circumnuclear material, driving powerful, large-scale outflows. Mrk~34 exemplifies this phase: its [O~III] emission extends in a bicone to $R_{\rm [O~III]}=2.2$~kpc, showing broad, high-velocity components ($v\sim1100$~km~s$^{-1}$, FWHM~$\sim$1700~km~s$^{-1}$; \citealt{fischer_hubble_2018}). It hosts the highest [O~III] mass outflow rate and kinetic luminosity in the sample, suggesting that the dense circumnuclear medium has been largely expelled, allowing the AGN to efficiently couple its energy to the host ISM and sustain galaxy-scale feedback \citep{trindade_falcao_hubble_2021-1}.
\end{description}

\noindent
Taken together, these stages outline a coherent evolutionary pathway for quasars. The sequence from pre-blowout to late post-blowout phases reflects the gradual coupling of the X-ray and [O~III] ionized gas phases as the AGN clears its environment, transitioning from compact, enshrouded systems to those exhibiting large-scale, energetically dominant outflows. A subsequent stage in this sequence may be represented by (unobscured) type~1 quasars (QSO1s), in which the circumnuclear gas has been completely cleared, giving the AGN a “naked” appearance, an unobscured view of the central engine once feedback has effectively evacuated the surrounding medium.

Supporting this scenario, \citet{trindade_falcao_hubble_2024} found that the sizes of [O~III]-emitting NLRs in type~1 AGNs (spanning $39.5<\log L_{\rm [O~III]}<43.3$~erg~s$^{-1}$) scale with luminosity as $\log R_{\mathrm{[O~III]},1}\propto(0.57\pm0.05)\log L_{\mathrm{[O~III]}}$, a significantly steeper relation than that found for type~2 AGNs over the same luminosity range, $\log R_{\mathrm{[O~III]},2}\propto(0.48\pm0.02)\log L_{\mathrm{[O~III]}}$. The combined type~1 + type~2 population follows an intermediate slope of $\log R_{\mathrm{[O~III]},1,2}\propto(0.51\pm0.03)\log L_{\mathrm{[O~III]}}$. For the subset of seven QSO1s, \citet{trindade_falcao_hubble_2024} additionally found significantly extended radial mass profiles, with the bulk of the [O~III]-emitting gas located at radii $>2$~kpc. Relative to the twelve QSO2s in \citet{fischer_hubble_2018}, these QSO1s have NLRs larger by roughly a factor of three, yet contain smaller total [O~III] gas masses, consistent with systems in which feedback has already cleared much of the circumnuclear medium and redistributed the remaining ionized gas to larger scales.

Future analysis of [O~III] kinematics in these QSO1s (\textit{Trindade Falcão et al., in prep.}) will further constrain how feedback operates in this unobscured phase.

\begin{table*}[t]
\centering
\caption{Evolutionary Stage Classification of the QSO2s from \citet{fischer_hubble_2018}.}
\label{tab:evol_sequence}
\begin{tabular}{cccccccc}
\hline\hline
\textbf{Object} & [O\,III] & $R_{\rm [O~III]}$ & $R_{\rm out}$ &
$v_{\rm max}$ & $R_{\rm dist}$ & FWHM$_{\rm nuclear}$ & Stage \\
&Morphology & (kpc) & (kpc) & (km~s$^{-1}$) & (kpc) & (km~s$^{-1}$) & \\
\hline
2MASX~J07594101  & Extended, conical & 3.2 & 0.67 & 315  & 0.67 & 1680 & II -- Blowout \\

2MASX~J08025293 & Extended, biconical  & 3.6 & 0.57 & 700 &0.89 & 870  & II -- Blowout \\

Mrk~34  & Extended, biconical & 2.2 & 1.89 & 1100 & 1.89& 580  & IV -- Late Post-Blowout \\

2MASX~J11001238 & Compact & 3.0 & 0.68 & 350 & $>$1.51 & 1780 & I -- Pre-Blowout \\

SDSS~J115245  & Extended, biconical & 3.6 & 0.15 & 300 & 1.23 & 360  & II -- Blowout \\

\textbf{FIRST~J120041}   & Extended, biconical  & 6.1 & 1.07 & 450  & $>$1.59 & 720  & III -- Early Post-Blowout \\

\textbf{2MASX~J130038} & Extended, biconical & 4.7 & 0.16 & 100  & 0.16 & 580  & II -- Blowout \\

2MASX~J14054117 & Compact  & 1.0 & 0.33 & 100 & $>$0.94 & 760  & I -- Pre-Blowout \\

B2~1435  & Extended, biconical & 3.8 & 0.20 & 200  & $>$1.74 & 630  & II -- Blowout \\

Mrk~477    & Extended, conical  & 2.7 & 0.54 & 500 & 0.90 & 2040 & II -- Blowout \\

2MASX~J16531506 & Extended, biconical & 6.50 & 0.57 & 110 & 1.23 & 510  & II -- Blowout \\

2MASX~J17135038 & Compact  & 1.8 & 0.65 & 160 & $>$0.92 & 1660 & I -- Pre-Blowout \\
\hline
\end{tabular}
\tablecomments{Morphological and kinematic parameters are from \citet{fischer_hubble_2018},  with evolutionary stage assignments following the framework defined in Section~\ref{sec:evolution}. $R_{\rm [O~III]}$ is the maximum deprojected NLR extent, $R_{\rm out}$ is the maximum deprojected distance of outflowing kinematics, and $R_{\rm dist}$ is the maximum projected distance of disturbed kinematics, for gas exhibiting FWHM$>$250~km~s$^{-1}$ that is at systemic or follows rotation. $v_{\rm max}$ is the maximum projected velocity measured inside $R_{\rm out}$ as the maximum outflow velocity. In boldface, are the two objects targeted in this work.}
\end{table*}


\section{Summary and Conclusions}
\label{sec:conclusions}

We present new Chandra/ACIS-S imaging spectroscopy of two luminous QSO2s from the \citet{fischer_hubble_2018} sample, FIRST~J120041.4+314745 and 2MASX~J13003807+5454367, and compare their X-ray properties with HST [O~III] morphologies and kinematics. Our analysis combines PSF modeling and subtraction (Section~\ref{sec:psf_model}), broad- and narrow-band imaging (Section~\ref{sec:imaging}), phenomenological spectral modeling (Section~\ref{sec:pheno}), and physically motivated photoionization modeling using \texttt{Cloudy} (Section~\ref{sec:physical}, and Section~\ref{sec:mod_xray}). The main results are as follows:

\begin{description}
    \item[~~1]\textbf{Both quasars show extended X-ray emission, but with different levels of correlation to the optical phase.} In FIRST~J120041, the X-rays form clumpy, kiloparsec-scale structures that closely trace the [O~III] bicone (Figure~\ref{fig:hst_FIRST}); peaks in soft X-ray surface brightness coincide with the regions of highest [O~III] velocities ($\sim$600-750~km~s$^{-1}$) and broadest line widths (up to $\sim$1700~km~s$^{-1}$; Figure~\ref{fig:hst_FIRST_slit}). In contrast, 2MASX~J130038 shows centrally concentrated X-ray emission with weak morphological correspondence to the [O~III] morphology (Figure~\ref{fig:hst_2MASX}) and no clear spatial correlation with the largely rotational [O~III] kinematics (Figure~\ref{fig:hst_2MASX_slit}).

    \item[~~2]\textbf{Photoionization dominates the soft X-ray emission in both quasars.} For FIRST~J120041, two photoionized components ($\log U$~=~0.5 and~2.5; Table~\ref{tab:spec_results}) reproduce the observed spectrum without requiring a thermal plasma. For 2MASX~J130038, a single medium ionization component ($\log U$~=~1) combined with a hard reprocessed continuum modeled with \texttt{MYTorus} provides a good fit (Figure~\ref{fig:photo_thermal_composite}). Adding a thermal component does not improve the fit quality in either case, implying that the soft X-ray emission is dominated by photoionized gas.

    \item[~~3]\textbf{The inferred hot-gas reservoirs are massive.} Within $r=1''$ apertures (2~kpc for FIRST~J120041 and 1.6~kpc for 2MASX~J130038), we estimate $M_{\rm x-ray}\sim4.1\times10^{8}~M_{\odot}$ and $M_{\rm x-ray}\sim~1.8\times10^{8}~M_{\odot}$, respectively, based on photoionization modeling (Table~\ref{tab:x-ray_masses}). Relative to the outflowing [O~III] gas \citep{trindade_falcao_hubble_2021-1}, the X-ray phase is roughly 4~times more massive (volume-normalized) in FIRST~J120041 and $\sim$16~times more massive in 2MASX~J130038. These results indicate that the hot, photoionized phase can dominate both the mass and energy budgets of the ionized ISM on kiloparsec scales.

    \item[~~4]\textbf{2MASX~J130038 shows a marginal evidence high-energy emission feature consistent with a nuclear hot wind.} An emission feature at $E_{\rm rest}=7.14^{+0.06}_{-0.06}$~keV improves the spectral fit by $\Delta$Cstat~=~10 for two additional free parameters, corresponding to a $\sim$2.7$\sigma$ detection. If interpreted as blueshifted Fe~XXVI Ly$\alpha$ ($v\sim7600$~km~s$^{-1}$), this feature would be consistent with highly ionized gas in the immediate nuclear region. Given its modest statistical significance, we regard this feature as suggestive rather than conclusive evidence for a nascent hot wind (Section~\ref{sec:ufo_emission}).

    \item[~~5]\textbf{A continuous evolutionary sequence may connect the two quasars and the broader QSO2 population.} Integrating the X-ray and optical morphologies, kinematics, and gas masses, we propose that luminous type~2 quasars evolve along a continuum sequence regulated by the gradual clearing of circumnuclear material by AGN radiation and winds (Section~\ref{sec:evolution}; Figure~\ref{fig:comparison}). In this framework, 2MASX~J130038 corresponds to an earlier phase where a weakly coupled, emerging X-ray wind begins to interact with the ISM (analogous to a “blowout” configuration), while FIRST~J120041 represents a relatively more developed stage where the hot wind actively entrains and drives the [O~III] gas (an “early post-blowout” configuration). More compact, enshrouded systems such as 2MASX~J11001238 or 2MASX~J17135038 may represent the most buried, confined phase of this sequence, “pre-blowout”, whereas sources like Mrk~34 illustrate the opposite extreme, a “late post-blowout” characterized by powerful, large-scale outflows and efficient AGN-ISM coupling (Table~\ref{tab:evol_sequence}).
\end{description}

Taken together, these results support a scenario in which hot, photoionized X-ray winds may act as the primary drivers of large-scale ionized outflows in luminous type~2 quasars. Where soft X-rays and [O~III] emission are morphologically and kinematically correlated (as in FIRST~J120041), the winds may be dynamically coupled to the host ISM; where the correlation is weak (as in 2MASX~J130038), the wind remains confined and has yet to fully disrupt the circumnuclear gas. 

The large inferred hot-gas masses, orders of magnitude above those found in nearby Seyferts such as NGC~1068 and NGC~4151, suggest that the X-ray phase may carry most of the energy and momentum available for feedback on kiloparsec scales. 

Finally, the broader distribution of the \citet{fischer_hubble_2018} QSO2 sample across this sequence of clearing and coupling implies that luminous obscured quasars may represent different points along a single evolutionary sequence driven by AGN feedback, likely culminating in unobscured type~1 quasars \citep{trindade_falcao_hubble_2024}, once the circumnuclear medium has been fully dispersed and the central engine is directly exposed.

\begin{acknowledgments}
We thank the anonymous referee for helpful comments that improved the clarity of this paper. ATF was supported by an appointment to the NASA Postdoctoral Program at the NASA Goddard Space Flight Center, administered by Oak Ridge Associated Universities under contract with NASA. LF was partially supported by Chandra/SAO grant GO3-24085X. TSB was supported by the Brazilian funding agency Conselho Nacional de Desenvolvimento Científico e Tecnológico (CNPq). M.V. gratefully acknowledges financial support from the Independent Research Fund Denmark via grant number DFF 3103-00146 and from the Carlsberg Foundation (grant CF23-0417). LCH was supported by the National Science Foundation of China (11991052, 12011540375, 12233001), the National Key R\&D Program of China (2022YFF0503401), and the China Manned Space Project (CMS-CSST-2021-A04, CMS-CSST-2021-A06).

This research has made use of the NASA/IPAC Extragalactic Database (NED), which is operated by the Jet Propulsion Laboratory, California Institute of Technology, under contract with the National Aeronautics and Space Administration. This paper used the photoionization code \texttt{CLOUDY}, which can be obtained from \url{http://www.nublado.org}.
\end{acknowledgments}

\newpage
\bibliography{references}{}
\bibliographystyle{aasjournalv7}


\end{document}